\documentclass[12pt]{iopart}

\usepackage[applemac]{inputenc}

\usepackage{iopams}
\usepackage{amssymb,amsfonts}
\usepackage{graphicx}
\usepackage{comment}

\usepackage{url}

\begin{document}

\title{Derivation of a Matrix-valued Boltzmann Equation for the Hubbard Model}

\author{Martin L.R. F\"urst$^{1,2,\mathrm{a}}$, Jani Lukkarinen$^{3,\mathrm{b}}$, Peng Mei$^{3,\mathrm{c}}$, Herbert Spohn$^{1,4,\mathrm{d}}$}

\address{$^1$ Mathematics Department, Technische Universit\"at M\"unchen, Boltzmannstra{\ss}e 3, 85748 Garching, Germany}
\address{$^2$ Excellence Cluster Universe, Technische Universit\"at M\"unchen, Boltzmannstra{\ss}e 2, 85748 Garching bei M\"unchen, Germany}
\address{$^3$ Department of Mathematics and Statistics, University of Helsinki, FI-00014 Helsingin yliopisto, Finland}
\address{$^4$ Physics Department, Technische Universit\"at M\"unchen, James-Franck-Stra{\ss}e 1, 85748 Garching bei M\"unchen, Germany}

\eads{$^\mathrm{a}$\mailto{mfuerst@ma.tum.de}, $^\mathrm{b}$\mailto{jani.lukkarinen@helsinki.fi}, $^\mathrm{c}$\mailto{peng.mei@helsinki.fi}, $^\mathrm{d}$\mailto{spohn@ma.tum.de}}

\begin{abstract}
For the spin-$\case{1}{2}$ Fermi-Hubbard model we derive the kinetic equation valid for weak interactions by using time-dependent perturbation expansion up to second order. In recent theoretical and numerical studies the kinetic equation has been merely stated without further details. In this contribution we provide the required background material.
\end{abstract}



%
%
%
%
%
%
\section{Introduction and main result} \label{sec:intro}

In its original form the Hubbard model is a simplified description of electrons in a solid, for which the lattice periodic potential is treated in the tight binding approximation and the interaction between electrons is reduced to an on-site potential. One particular realization would be graphene, which has stirred a lot of activity \cite{GN07,PAB04}. 
In graphene the ${\rm C}$ atoms form a sheet arranged as a honeycomb lattice resulting in a two-band Hubbard model. The energy bands exhibit conical intersections which are at the core of interesting dynamical behavior. The Fermi-Hubbard model has also been realized in a very different context as an accurate description of the motion of cold atoms in an optical lattice \cite{SHetal12}. Thereby one has at disposal new possibilities and methods to study the dynamical properties of the Hubbard model.

In our contribution, we will derive the kinetic equation for the Hubbard model which will be an accurate description for sufficiently small interactions. Let us first recall the structure of the Hubbard model. For simplicity, we restrict ourselves to the lattice $\mathbb{Z}^d$ (single band), but generalizations are easily implemented. The basic object is thus a spin-$\case{1}{2}$ Fermi field, $a_\sigma(x)$, $x \in \mathbb{Z}^d$, $\sigma \in \{ \uparrow, \downarrow\}$, with anticommutation relation,
\begin{eqnarray}
\{ a_\sigma(x)^*, a_\tau(y) \} = \delta_{xy} \delta_{\sigma \tau}, \\ 
\{ a_\sigma(x), a_\tau(y) \} =  0, \\
\{ a_\sigma(x)^*, a_\tau(y)^* \} = 0,
\end{eqnarray}
where $\{ A, B \} = A B + B A$ and $A^*$ denotes the adjoint operator to $A$. The Hubbard hamiltonian reads
\begin{eqnarray}
\fl H = \sum_{x, y \in \mathbb{Z}^d} \alpha(x - y) \, a(x)^* \cdot a(y)
		+ \lambda \, \case{1}{2} \sum_{x,y \in \mathbb{Z}^d} V(x-y) \big( a(x)^* \cdot a(x) \big) \big( a(y)^* \cdot a(y) \big)
\end{eqnarray}
with $a(x)^* \cdot a(y) = \sum_{ \sigma \in \{ \uparrow, \downarrow\} } a_\sigma(x)^* \, a_\sigma(y)$. $\alpha$ is the hopping amplitude, with the properties $\alpha(x) = \overline{\alpha}(x)$, $\alpha(x) = \alpha(-x)$. The particles interact through the weak pair potential $\lambda V$, $V: \mathbb{Z}^d \rightarrow \mathbb{R}$, $V(x) = V(-x)$. $V$ is assumed to decay fast enough to be absolutely summable and $0 < \lambda \ll 1$. A particular case of interest is the on-site, $\delta$-potential $V(x) = \delta_{0x}$. Our notation emphasizes the invariance under global spin rotations.

For the Fourier transformation we use the convention
\begin{equation} \label{eq:Fouriertransformation}
\hat{f}(k) = \sum_{x \in \mathbb{Z}^d} f(x)\, \mathrm{e}^{-2 \pi \mathrm{i} \, k \cdot x}
\end{equation}
and correspondingly for operator valued functions; for instance,
$\{\hat{a}_\sigma(k)^*,\hat{a}_\tau(\tilde{k})\} = \delta_{\sigma\tau} \, \delta(k-\tilde{k})$.
Then the first Brillouin zone is the set $\mathbb{T}^{d} = [-\case{1}{2}, \case{1}{2}]^{d}$ with periodic boundary conditions. The dispersion relation is $\omega(k) = \hat{\alpha}(k)$ and in Fourier space $H$ can be written as
\begin{eqnarray} \label{eq:HubbardHamiltonian}
\fl H = H_0 + \lambda H_1 = \int_{\mathbb{T}^{d}} \mathrm{d} k \, \omega(k) \, \hat{a}(k)^* \cdot \hat{a}(k) \nonumber \\ 
\fl \quad + \lambda \, \case{1}{2} \int_{(\mathbb{T}^{d})^4} \mathrm{d} k_{1234} \, \delta(k_1 - k_2 + k_3 - k_4) \hat{V}(k_1-k_2)
  \big( \hat{a}(k_1)^* \cdot \hat{a}(k_2) \big) \, \big( \hat{a}(k_3)^* \cdot \hat{a}(k_4) \big)\, ,
\end{eqnarray}
where $\mathrm{d} k_{1234} = \mathrm{d} k_1\, \mathrm{d} k_2\, \mathrm{d} k_3\, \mathrm{d} k_4$.

In the spatially homogeneous case the central quantity is the time dependent average Wigner matrix $W$ defined by
\begin{equation} \label{eq:InitialCondition}
\langle \hat{a}_\sigma(k,t)^* \hat{a}_\tau(\tilde{k},t) \rangle = \delta(k - \tilde{k}) W_{\sigma \tau}(k,t),
\end{equation}
which for times up to order $\lambda^{-2}$ will satisfy in approximation a kinetic equation. In~(\ref{eq:InitialCondition}), $\langle \cdot \rangle$ denotes the average over the initial state and the operators are computed in the Heisenberg picture, $A(t) = \mathrm{e}^{\mathrm{i} H t} A \mathrm{e}^{-\mathrm{i} H t}$. 

Our goal is to derive the Boltzmann type equation for the hamiltonian (\ref{eq:HubbardHamiltonian}). At first sight this may look like a problem treated in textbooks. However, to the best of our knowledge, the standard discussion assumes that initially $W_{\sigma \tau}(k) = \delta_{\sigma \tau} W_\sigma(k)$, a property which is preserved by the kinetic equation. One obtains then a coupled set of equations for $W_\uparrow$ and $W_\downarrow$, which have the same structure as a kinetic equation for a classical two-component system. Physically, there is no compelling reason to have $W$ diagonal. In fact, interesting aspects of the spin dynamics may be lost. In our contribution we will treat general initial Wigner matrices. This may look like an easy excerise---now there will simply be a coupled set of equations for $W_{\uparrow \uparrow}$, $W_{\uparrow \downarrow}$, $W_{\downarrow \downarrow}$. However, already at second order, the number of terms steeply increases. Even more importantly, one looses sight of any 
comprehensible structure. A more optimal strategy is to regard $W(k,t)$ as a $2 \times 2$ matrix and to completely avoid the representation in a specific spin basis. As a result our kinetic equation has the familiar structure, except for a particular ordering of products of $W$'s and not necessarily expected trace terms.

The Hubbard hamiltonian (\ref{eq:HubbardHamiltonian}) has a twofold degenerate band.  Compared to the earlier studied cases, such as the closely related derivation of the fermionic Boltzmann-Nordheim equation without spin \cite{LS09b}, this leads to a further novel feature of the kinetic equation. Besides the entropy generating collision term, there is a Vlasov type term of the form $- \mathrm{i} [H_\mathrm{eff}(k,t), W(k,t)]$, where $H_\mathrm{eff}(k,t)$ is a $2 \times 2$ matrix which depends itself quadratically on $W(t)$. In addition the effective hamiltonian carries the contribution
\begin{equation}
\lambda^{-1} \int_{\mathbb{T}^{d}} \mathrm{d} \tilde{k} \, \hat{V}(k - \tilde{k}) W(\tilde{k},t)
\end{equation}
which generates fast oscillations on the kinetic time scale. Of course, such a unitary evolution does not produce any entropy.

As our main result we obtain a kinetic equation, valid for the kinetic time scale where $t$ is replaced by $\lambda^{-2} t$. It is an evolution equation for the $2 \times 2$ matrix-valued Wigner function of the form
\begin{equation} \label{eq:BoltzmannEquation:1}
\frac{\partial}{\partial t} W(t) = \mathcal{C}[W(t)], \qquad \qquad 
	\mathcal{C}[W] = \mathcal{C}_\mathrm{c}[W] + \mathcal{C}_\mathrm{d}[W],
\end{equation}
which has to be supplemented with some initial condition $W(k,0) = W^{(0)}(k)$. The conservative term, $\mathcal{C}_c$, has the form
\begin{equation} \label{eq:Cc:1}
\fl \mathcal{C}_\mathrm{c}[W(t)](k) = - \mathrm{i} \, [H_\mathrm{eff}[W(t)](k), W(k,t)] - \lambda^{-1} \mathrm{i} [R[W(t)](k), W(k,t)].
\end{equation}
Here the effective hamiltonian is given by
\begin{eqnarray} \label{eq:Heff:1}
\fl H_{\mathrm{eff}}[W]_1 = \int_{(\mathbb{T}^{d})^3} \mathrm{d} k_{234} \, \delta(k_1 + k_2 - k_3 - k_4) \, 
	\mathcal{P} \left( \case{1}{\omega_1 + \omega_2 - \omega_3 - \omega_4} \right) \nonumber \\
	\times \big( \hat{V}(k_2-k_3) \hat{V}(k_2-k_4) \mathcal{A}_\mathrm{X,c}[W]_{234} + \hat{V}(k_2-k_4)^2 \mathcal{A}_\mathrm{tr,c}[W]_{234} \big),
\end{eqnarray}
where
\begin{eqnarray}
\mathcal{A}_\mathrm{X,c}[W]_{234} = W_4 W_3 - W_2 W_4 - W_4 W_2 + W_2
\end{eqnarray}
and
\begin{eqnarray}
\mathcal{A}_\mathrm{tr,c}[W]_{234} = \big( \mathrm{tr}[W_2] - \mathrm{tr}[W_4] \big) W_3,
\end{eqnarray}
with $\mathrm{d} k_{234} = \mathrm{d} k_2 \mathrm{d} k_3 \mathrm{d} k_4$. We have introduced the shorthand notations $W_j = W(k_j,t)$, $\omega_j = \omega(k_j)$, $H_{\mathrm{eff},1} = H_\mathrm{eff}(k_1,t)$. Since $W$ is $2 \times 2$ matrix-valued, $\mathrm{tr}[\,\cdot\,]$ is the trace in spin space. Finally, $\mathcal{P}$ denotes a principal value integral: the notation $\mathcal{P}(1/f(k))$ means that for small $\varepsilon>0$, we first integrate over $k$ with $|f(k)|>\varepsilon$, and the result is then given by the $\varepsilon\to 0$ limit of these integrals.  Since the $k_3$, $k_4$ integration can be interchanged, $H_{\mathrm{eff}} = H_{\mathrm{eff}}^*$, as it should be. The second summand in (\ref{eq:Cc:1}) is linear in $W$ and reads
\begin{equation}
R[W](k) = \int_{\mathbb{T}^{d}} \mathrm{d} \tilde{k} \, \hat{V}(k - \tilde{k}) W(\tilde{k}).	
\end{equation}
The dissipative part of the collision term, $\mathcal{C}_\mathrm{d}$, is given by
%
%
%
\begin{eqnarray} \label{diss:jan}
\fl \mathcal{C}_\mathrm{d}[W]_1 = \pi \int_{(\mathbb{T}^{d})^3} \mathrm{d} k_{234} \, \delta(k_1 + k_2 - k_3 - k_4) 
		\delta(\omega_1 + \omega_2 - \omega_3 - \omega_4) \nonumber \\
 	\fl \qquad \times \Big( \hat{V}(k_2 - k_3) \hat{V}(k_2 - k_4) \mathcal{A}_\mathrm{quad,d}[W]_{1234} + \hat{V}(k_2 - k_4)^2 \mathcal{A}_\mathrm{tr,d}[W]_{1234} \Big),
\end{eqnarray}
where
\begin{eqnarray}
\fl \mathcal{A}_\mathrm{quad,d}[W]_{1234} = \tilde{W}_4 W_2 \tilde{W}_3 W_1 - W_4 \tilde{W}_2 W_3 \tilde{W}_1 - \tilde{W}_1 W_3 \tilde{W}_2 W_4 + W_1 \tilde{W}_3 W_2 \tilde{W}_4
\end{eqnarray}
and
\begin{eqnarray}
\fl \mathcal{A}_\mathrm{tr,d}[W]_{1234} = \big( \tilde{W}_1 W_3 + W_3 \tilde{W}_1 \big) \, \mathrm{tr}[\tilde{W}_2 W_4] - \big( W_1 \tilde{W}_3 + \tilde{W}_3 W_1 \big) \, \mathrm{tr}[W_2 \tilde{W}_4]
\end{eqnarray}
with $\tilde{W} = 1_{\mathbb{C}^2} - W$. Note that in the commuting scalar case $H_\mathrm{eff}$ and $R$ have no effect.

In the special case of an on-site potential, one finds $\hat{V} = 1$ and hence $R[W](k) = \int_{\mathbb{T}^d} \mathrm{d} \tilde{k} W(\tilde{k})$. $R[W]$ does not depend on $k$ and, by conservation of spin, also not on $t$. We denote the resulting matrix by $R$, which still depends on the initial condition. Since $R$ is constant, it can be removed explicitly through the unitary transformation
\begin{equation}
W(k,t) \mapsto \mathrm{e}^{\mathrm{i} \lambda^{-1} R t} \, W(k,t) \, \mathrm{e}^{-\mathrm{i} \lambda^{-1} R t}.
\end{equation}
For a general potential, one has to solve the full nonlinear equation including the $R[W(t)]$ term.
\section{Expansion in $\lambda$} \label{section:2}
We assume that the initial state, $\langle \cdot \rangle$, is gauge invariant, invariant under translations, and quasi-free. The state is then completely determined by the two point function 
\begin{eqnarray}
\langle \hat{a}_\sigma(k)^* \hat{a}_\tau(\tilde{k}) \rangle = \delta(k - \tilde{k}) W_{\sigma \tau}(k,0), \quad \sigma, \tau \in \{\uparrow, \downarrow \}.	
\end{eqnarray}
Averages of the form $\langle (a^*)^m a^n \rangle$ vanish unless $m = n$ and all other moments are determined by the Wick pairing rule. The state at time $t$ will still be gauge invariant and invariant under translations. But quasi-freeness will not be preserved. The basic tenet of kinetic theory is that for small $\lambda$ and times of order $\lambda^{-2}$ the quasi-free property is approximately maintained. However, the initial $W(0)$ will have evolved to $W(t)$ (on the time scale $\lambda^{-2}$). To determine the collision operator $\mathcal{C}$ of (\ref{eq:BoltzmannEquation:1}) one has to study the increment $W(t + \mathrm{d}t) - W(t)$. Since $W(t)$ is approximately quasi-free by assumption, we might as well set $t = 0$ and then evaluate the collision operator at $W(\mathrm{d}t)$. (This is a version of the much debated repeated random phase approximation.) $\mathrm{d}t$ is long on the microscopic scale, but short on the kinetic scale. 

More formally, we expand the true two-point function $W_\lambda$, defined by 
the relation $\delta(k - \tilde{k}) W_\lambda(k,t)_{\sigma \tau} = \langle \hat{a}_\sigma(k,t)^* \hat{a}_\tau(\tilde{k},t) \rangle$,
for fixed $t$ up to order $\lambda^{2}$ as
\begin{equation}
W_\lambda(k,t) = W^{(0)}(k) + \lambda W^{(1)}(k,t) + \lambda^2 W^{(2)}(k,t) + \mathcal{O}(\lambda^3).	
\end{equation}
The collision operator will then be extracted from $W^{(2)}$, see Section \ref{section3}. But the main effort is to properly organize the expansion. To avoid a specific spin basis, we choose arbitrary vectors  $\mathrm{f}, \mathrm{g} \in \mathbb{C}^2$ 
and consider  $\langle \mathrm{f}, \, W_\lambda(k,t) \mathrm{g} \rangle$ where $\langle \cdot, \, \cdot \rangle$ denotes the inner product (anti-linear on the left) in spin space. It is advantageous to 
introduce the vector valued operators
\begin{equation} \label{op:vec}
\hat{a}_\mathrm{f}(k)^* = \left( \begin{array}{c} \overline{\mathrm{f}}_\uparrow \, \hat{a}_\uparrow(k)^* \\ \overline{\mathrm{f}}_\downarrow \, \hat{a}_\downarrow(k)^* \end{array} \right) \qquad \mathrm{and} \qquad
\hat{a}_\mathrm{g}(k) = \left( \begin{array}{c} \mathrm{g}_\uparrow \, \hat{a}_\uparrow(k) \\ \mathrm{g}_\downarrow \, \hat{a}_\downarrow(k) \end{array} \right), 
\end{equation}
where $\overline{\mathrm{f}}$ denotes complex conjugate and $\mathrm{f}_\sigma, \mathrm{g}_\sigma$, $\sigma \in \{\uparrow, \downarrow\}$ denote the components of $\mathrm{f}$ and $\mathrm{g}$.  We will also use the following operations mapping two $2$-vector valued operators into a scalar-valued one:
\begin{equation}
v \odot w = \sum_{\sigma, \tau \in \{\uparrow, \downarrow \} } v_\sigma w_\tau \qquad \mathrm{and} \qquad v \cdot w = \sum_{\sigma \in \{\uparrow, \downarrow \} } v_\sigma w_\sigma.	
\end{equation}
For instance, then $\langle \hat{a}_\mathrm{f}(k,t)^* \odot \hat{a}_\mathrm{g}(\tilde{k},t) \rangle = \delta(k - \tilde{k}) \, \langle \mathrm{f}, \, W_\lambda(k,t) \mathrm{g} \rangle$.

Let us first compute the time derivative of the basic $2$-vector valued operator
\begin{eqnarray}
\frac{\mathrm{d}}{\mathrm{d}t} \hat{a}_\mathrm{f}(k,t)^{\#} = \mathrm{i} [H, \hat{a}_\mathrm{f}(k,t)^{\#}] 
= \mathrm{i} [H_0, \hat{a}_\mathrm{f}(k)^{\#}](t) + \lambda \mathrm{i} \, [H_1, \hat{a}_\mathrm{f}(k)^{\#}](t),
\end{eqnarray}
where $\#$ denotes either nothing or an adjoint (annihilation or creation operator). For the quadratic $H_0$ it follows directly from the commutation relations that
\begin{equation}
[H_0, \hat{a}_\mathrm{g}(k)] = \int_{\mathbb{T}^{d}} \mathrm{d} k' \, \omega(k') [\hat{a}(k')^* \cdot \hat{a}(k'), 
	\hat{a}_\mathrm{g}(k)] = - \omega(k) \, \hat{a}_\mathrm{g}(k),
\end{equation}
and
\begin{equation}
[H_0, \hat{a}_\mathrm{f}(k)^*] = - [H_0, \hat{a}_\mathrm{f}(k)]^* = \omega(k) \, \hat{a}_\mathrm{f}(k)^*.	
\end{equation}
For $H_1$ we use
\begin{eqnarray} 
\fl [H_1, \hat{a}_\mathrm{g}(k)] = \case{1}{2} \int_{(\mathbb{T}^{d})^4} \mathrm{d} k_{1234} \, \delta(\underline{k}) \, \hat{V}(k_2 - k_3) \,
	[ \big( \hat{a}(k_1)^* \cdot \hat{a}(k_2) \big) \, \big( \hat{a}(k_3)^* \cdot \hat{a}(k_4) \big), \, \hat{a}_\mathrm{g}(k) ], 	
\end{eqnarray}
with $\underline{k} = k_1 - k_2 + k_3 - k_4$. Using the commutators
\begin{eqnarray} \label{comm}
\fl [ \big( \hat{a}(k_1)^* \cdot \hat{a}(k_2) \big) \, \big( \hat{a}(k_3)^* \cdot \hat{a}(k_4) \big), \, \hat{a}_\mathrm{g}(k) ] =
	 - \delta(k_1 - k) \, \hat{a}_\mathrm{g}(k_2) \, \big( \hat{a}(k_3)^* \cdot \hat{a}(k_4) \big) \nonumber \\
	 - \delta(k_3 - k) \, \hat{a}_\mathrm{g}(k_4) \, \big( \hat{a}(k_1)^* \cdot \hat{a}(k_2) \big)
	 + \delta(k_1 - k_4) \delta(k_3 - k) \, \hat{a}_\mathrm{g}(k_2),
\end{eqnarray}
we obtain
\begin{eqnarray} \label{evol}
\fl \frac{\mathrm{d}}{\mathrm{d}t} \hat{a}_\mathrm{g}(k_1,t) = \mathrm{i} [H, \hat{a}_\mathrm{g}(k_1,t)]
	= - \mathrm{i} \, \omega(k_1) \, \hat{a}_\mathrm{g}(k_1,t) + \lambda \frac{\mathrm{i}}{2} \, V(0) \, \hat{a}_\mathrm{g}(k_1,t) \nonumber \\
- \mathrm{i} \lambda \int_{(\mathbb{T}^{d})^3} \mathrm{d} k_{234} \, \delta(\underline{k}) \, \hat{V}(k_3 - k_4) \,
	\hat{a}_\mathrm{g}(k_2,t) \, \big( \hat{a}(k_3,t)^* \cdot \hat{a}(k_4,t) \big).
\end{eqnarray}

To proceed further we need convenient shorthands. With the notation $k_{1234} = (k_1,k_2,k_3,k_4)$ and for complex-valued functions $h$, we set 
\begin{eqnarray}
\fl \mathcal{A}[h,a,b,c](k_1,t) = \int_{(\mathbb{T}^{d})^3} \mathrm{d} k_{234} \, \delta(\underline{k}) \, h(k_{1234},t) \, \hat{V}(k_3 - k_4) \,
	a(k_2,t) \, \big( b(k_3,t) \cdot c(k_4,t) \big), \\
\fl \mathcal{A}_*[\overline{h},a,b,c](k_1,t) = \int_{(\mathbb{T}^{d})^3} \mathrm{d} k_{234} \, \delta(\underline{k}) \, 
	\overline{h}(k_{1234},t) \, \hat{V}(k_2 - k_3) \, \big( a(k_2,t) \cdot b(k_3,t) \big) \, c(k_4,t),	
\end{eqnarray}
where $a, b, c$ are two-component vector-valued operators as in (\ref{op:vec}). Then $\mathcal{A}$ is again a vector-valued operator and it holds
\begin{equation}
\big( \mathcal{A}[h,a,b^*,c](k,t) \big)^* = \mathcal{A}_*[\overline{h},c^*,b,a^*](k,t). 	
\end{equation}
With this notation we have
\begin{eqnarray}
\fl \mathcal{A}[\mathrm{id},\hat{a}_\mathrm{g},\hat{a}^*,\hat{a}](k_1,t) 
	= \int_{(\mathbb{T}^{d})^3} \mathrm{d} k_{234} \, \delta(\underline{k}) \, \hat{V}(k_3 - k_4) \,
		\hat{a}_\mathrm{g}(k_2,t) \, \big( \hat{a}(k_3,t)^* \cdot \hat{a}(k_4,t) \big), \\
\fl \mathcal{A}_*[\mathrm{id},\hat{a}^*,\hat{a},\hat{a}^*_\mathrm{f}](k_1,t) 
	= \int_{(\mathbb{T}^{d})^3} \mathrm{d} k_{234} \, \delta(\underline{k}) \, \hat{V}(k_2 - k_3) \,
		\big( \hat{a}(k_2,t)^* \cdot \hat{a}(k_3,t) \big) \, \hat{a}_\mathrm{f}(k_4,t)^*.		
\end{eqnarray}
where ``$\mathrm{id}$'' is the identity function. The evolution equation (\ref{evol}) is then
\begin{equation}
\fl \frac{\mathrm{d}}{\mathrm{d}t} \hat{a}_\mathrm{g}(k,t) = - \mathrm{i} \left( \omega(k) - \case{\mathrm{1}}{2} \lambda V(0) \right)
	\hat{a}_\mathrm{g}(k,t) - \mathrm{i} \lambda \, \mathcal{A}[\mathrm{id},\hat{a}_\mathrm{g},\hat{a}^*,\hat{a}](k,t)
\end{equation}
and correspondingly for the creation operator
\begin{equation}
\fl \bigg( \frac{\mathrm{d}}{\mathrm{d}t} \hat{a}_\mathrm{f}(k,t) \bigg)^* = \frac{\mathrm{d}}{\mathrm{d}t} \hat{a}_\mathrm{f}(k,t)^*
	= \mathrm{i} \left( \omega(k) - \case{1}{2} \lambda V(0) \right) \, \hat{a}_\mathrm{f}(k,t)^* 
		+ \mathrm{i} \lambda \, \mathcal{A}_*[\mathrm{id},\hat{a}^*,\hat{a},\hat{a}^*_\mathrm{f}](k,t).
\end{equation}
The linear part can be removed through defining
\begin{equation}
	\mathfrak{a}_\mathrm{g}(k,t) = \mathrm{e}^{ \mathrm{i} ( \omega(k) - \case{1}{2} \lambda V(0) ) t} \, \hat{a}_\mathrm{g}(k,t). 	
\end{equation}
where $\mathfrak{a}$ always acts in Fourier space. Clearly,
\begin{equation}
\mathfrak{a}_\mathrm{f}^*(k,t) = \big( \mathfrak{a}_\mathrm{f}(k,t) \big)^* = \mathrm{e}^{-\mathrm{i} ( \omega(k) - \case{1}{2} \lambda V(0) ) t} \, \hat{a}_\mathrm{f}(k,t)^*	
\end{equation}
and for the correlation it still holds that
\begin{equation}
\langle \mathfrak{a}^*_\mathrm{f}(k,t) \odot \mathfrak{a}_\mathrm{g}(\tilde{k},t) \rangle
	= \langle \hat{a}_\mathrm{f}(k,t)^* \odot \hat{a}_\mathrm{g}(\tilde{k},t) \rangle.      	
\end{equation}
Introducing the further shorthand
\begin{equation}
\omega_{abcd} = \omega(k_a) - \omega(k_b) + \omega(k_c) - \omega(k_d)    
\end{equation}
one finally arrives at
\begin{equation} \label{time:iter}
\frac{\mathrm{d}}{\mathrm{d}t} \mathfrak{a}_\mathrm{g}(k_1,t)
	= - \mathrm{i} \lambda \, \mathcal{A}[\mathrm{e}^{\mathrm{i} \omega_{1234} t}, \mathfrak{a}_\mathrm{g}, \mathfrak{a}^*,
		\mathfrak{a}](k_1,t)
\end{equation}
and for the adjoint
\begin{equation}
\frac{\mathrm{d}}{\mathrm{d}t} \mathfrak{a}_\mathrm{f}^*(k_1,t)
	= \mathrm{i} \lambda \, \mathcal{A}_*
		[\mathrm{e}^{-\mathrm{i} \omega_{1234} t}, \mathfrak{a}^*, \mathfrak{a}, \mathfrak{a}^*_\mathrm{f}](k_1,t).  
\end{equation}

By the fundamental theorem of calculus
\begin{equation}
\int_{0}^t \mathrm{d}s \frac{\mathrm{d}}{\mathrm{d}s} \mathfrak{a}^{\#}_\mathrm{f}(k,s) 
	= \mathfrak{a}^{\#}_\mathrm{f}(k,t) - \mathfrak{a}^{\#}_\mathrm{f}(k,0),	
\end{equation}
which implies
\begin{equation}
\mathfrak{a}_\mathrm{g}(k_1,t) = \mathfrak{a}_\mathrm{g}(k_1,0) 
	- \mathrm{i} \lambda \int_{0}^t \mathrm{d}s \, 
		\mathcal{A}[\mathrm{e}^{\mathrm{i} \omega_{1234} s}, \mathfrak{a}_\mathrm{g}, \mathfrak{a}^*, \mathfrak{a}](k_1,s). 
\end{equation}
Iterating (\ref{time:iter}) twice up to second order of the Dyson expansion, with an error of order~$\lambda^3$,
\begin{eqnarray}
\fl \frac{\mathrm{d}}{\mathrm{d}t} \mathfrak{a}_\mathrm{g}(k_1,t) = - \mathrm{i} \lambda \, \mathcal{A}[\mathrm{e}^{\mathrm{i} \omega_{1234} t}, \hat{a}_\mathrm{g}, \hat{a}^*, \hat{a}](k_1,0) 
\nonumber \\ \quad
- \lambda^2 \, \int_0^t \mathrm{d}s \,
		\mathcal{A}[\mathrm{e}^{\mathrm{i} \omega_{1234} t},
		\mathcal{A}[\mathrm{e}^{\mathrm{i} \omega_{2678} s}, \hat{a}_\mathrm{g}, \hat{a}^*, \hat{a}],
		\hat{a}^*, \hat{a}](k_1,s) 
\nonumber \\ \quad
+ \lambda^2 \, \int_0^t \mathrm{d}s \,
		\mathcal{A}[\mathrm{e}^{\mathrm{i} \omega_{1234} t}, \hat{a}_\mathrm{g}, 
		\mathcal{A}_*[\mathrm{e}^{-\mathrm{i} \omega_{3678} s}, \hat{a}^*, \hat{a}, \hat{a}^*], \hat{a}](k_1,s) 
\nonumber \\ \quad
- \lambda^2 \, \int_0^t \mathrm{d}s \,
		\mathcal{A}[\mathrm{e}^{\mathrm{i} \omega_{1234} t}, \hat{a}_\mathrm{g}, \hat{a}^*,
		\mathcal{A}[\mathrm{e}^{\mathrm{i} \omega_{4678} s}, \hat{a}, \hat{a}^*, \hat{a}]](k_1,s) + \mathcal{O}(\lambda^3) \nonumber \\
=	\lambda \, \frac{\mathrm{d}}{\mathrm{d}t} \mathfrak{a}_{\mathrm{g}}^{(1)}(k_1,t) 
		+ \lambda^2 \, \frac{\mathrm{d}}{\mathrm{d}t} \mathfrak{a}^{(2)}_{\mathrm{g}}(k_1,t) + \mathcal{O}(\lambda^3).
\end{eqnarray}
Hence, for fixed $t$, as an expansion in $\lambda$,
\begin{equation}
\mathfrak{a}_\mathrm{g}(k,t) = \mathfrak{a}^{(0)}_{\mathrm{g}}(k,t) + \lambda \, \mathfrak{a}^{(1)}_{\mathrm{g}}(k,t) 
	+ \lambda^2 \, \mathfrak{a}^{(2)}_{\mathrm{g}}(k,t) + \mathcal{O}(\lambda^3),
\end{equation}
where $\mathfrak{a}^{(0)}_{\mathrm{g}}(k,t) = \mathfrak{a}^{(0)}_{\mathrm{g}}(k,0) =\hat{a}_\mathrm{g}(k)$. A corresponding expression is satisfied by $\mathfrak{a}^*_{\mathrm{f}}(k,t)$. Iterating further yields the formal expansion
\begin{equation} \label{expansion}
\frac{\mathrm{d}}{\mathrm{d}t} \langle \mathfrak{a}^*_\mathrm{f}(k,t) \odot \mathfrak{a}_\mathrm{g}(\tilde{k},t) \rangle
	 = \sum_{n=0}^\infty \lambda^n \sum_{m=0}^n \frac{\mathrm{d}}{\mathrm{d}t} \langle \mathfrak{a}^*_\mathrm{f}(k,t)^{(m)} \odot \mathfrak{a}_\mathrm{g}(\tilde{k},t)^{(n-m)} \rangle.
\end{equation}
Therefore, $W_\lambda(k,t)$ can be written as
\begin{eqnarray} \label{expansion:2}
\fl \delta(k - \tilde{k}) \, \langle \mathrm{f}, \, W_\lambda(k,t) \mathrm{g} \rangle = 
	\langle \mathfrak{a}^*_\mathrm{f}(k,0) \odot \mathfrak{a}_\mathrm{g}(\tilde{k},0) \rangle 
\nonumber \\ \quad
+ \sum_{n=1}^\infty \lambda^n \int_0^t \mathrm{d}s \sum_{m=0}^n \frac{\mathrm{d}}{\mathrm{d}s} 
		\langle \mathfrak{a}^*_\mathrm{f}(k,s)^{(m)} \odot \mathfrak{a}_\mathrm{g}(\tilde{k},s)^{(n-m)} \rangle \nonumber \\
= \delta(k - \tilde{k}) \, \langle \mathrm{f}, \, W^{(0)}(k,t) \mathrm{g} \rangle 
	+ \delta(k - \tilde{k}) \sum_{n=1}^\infty \lambda^n \langle \mathrm{f}, \, W^{(n)}(k,t) \mathrm{g} \rangle.  
\end{eqnarray}
%
The zeroth order term of equation (\ref{expansion:2}) reads
\begin{equation}
\delta(k - \tilde{k}) \, \langle \mathrm{f}, \, W^{(0)}(k) \mathrm{g} \rangle 
	= \langle \mathfrak{a}^*_\mathrm{f}(k,0) \odot \mathfrak{a}_\mathrm{g}(\tilde{k},0) \rangle
	= \langle \hat{a}^*_\mathrm{f}(k) \odot \hat{a}_\mathrm{g}(\tilde{k}) \rangle.
\end{equation}
In the next two sections we determine the terms of first and second order.
\section{First order terms}
\begin{figure}
	\centering
	\includegraphics[width=0.9\columnwidth]{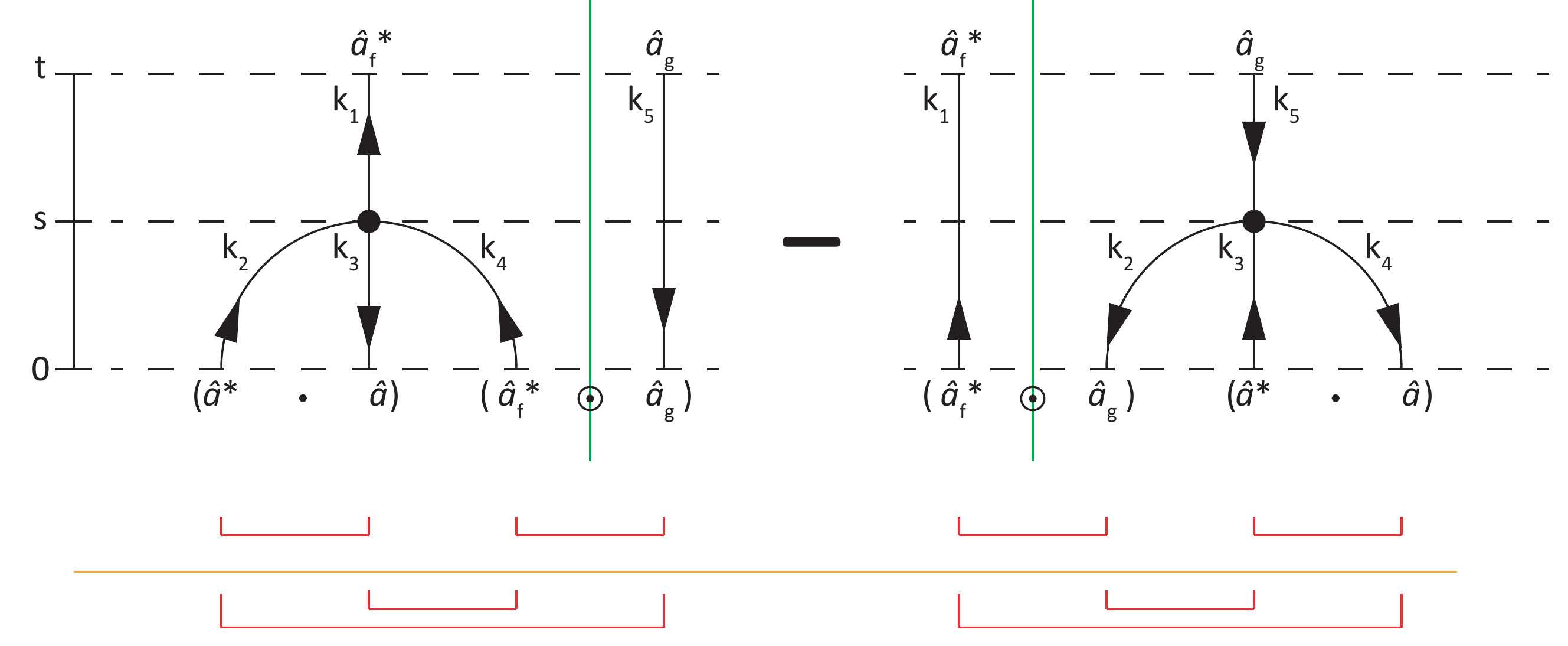}
	\caption{The full diagrams of the terms at first order in $\lambda$.} \label{firstOrder:graphs}
\end{figure}

Let us consider the $W^{(1)}(k,t)$-term of equation (\ref{expansion:2}). Its structure will be easier to capture once we represent the various summands as Feynman diagrams. The first order terms are determined by
\begin{eqnarray}\label{eq:full1storder}
\fl \delta(k_1 - k_5) \langle \mathrm{f}, \, W^{(1)}(k_1,t) \mathrm{g} \rangle \nonumber \\
\fl \qquad = \mathrm{i} \int_0^t \mathrm{d}s \, \langle \mathcal{A}_*[\mathrm{e}^{-\mathrm{i} \omega_{1234} s}, \mathfrak{a}^*, \mathfrak{a},
   \mathfrak{a}^*_\mathrm{f}](k_1) \odot \mathfrak{a}_{\mathrm{g}}(k_5,s)^{(0)} \rangle \nonumber \\
\fl \qquad \quad - \mathrm{i} \int_0^t \mathrm{d}s \, \langle \mathfrak{a}_{\mathrm{f}}^*(k_1,s)^{(0)} \odot \mathcal{A}[\mathrm{e}^{\mathrm{i} \omega_{5234} s}, \mathfrak{a}_\mathrm{g}, \mathfrak{a}^*, \mathfrak{a}](k_5) \rangle \nonumber \\
\fl \qquad = \mathrm{i} \int_0^t \mathrm{d}s \int_{(\mathbb{T}^{d})^3} \mathrm{d} k_{234} \, \delta(\underline{k}) \, \hat{V}(k_2-k_3) \,
		\mathrm{e}^{-\mathrm{i} \omega_{1234} s} \langle \big( \hat{a}(k_2)^* \cdot \hat{a}(k_3) \big) 
		\big(\hat{a}_\mathrm{f}(k_4)^* \odot \hat{a}_{\mathrm{g}}(k_5) \big) \rangle \nonumber \\
\fl \qquad - \mathrm{i} \int_0^t \mathrm{d}s \int_{(\mathbb{T}^{d})^3} \mathrm{d} k_{234} \, \delta(\underline{k}) \, \hat{V}(k_3 - k_4) \,
		\mathrm{e}^{\mathrm{i} \omega_{5234} s} \langle \big( \hat{a}_\mathrm{f}(k_1)^* \odot \hat{a}_{\mathrm{g}}(k_2) \big) 
		\big(\hat{a}(k_3)^* \cdot \hat{a}(k_4) \big) \rangle.	
\end{eqnarray}
where $\dot{\mathfrak{a}}(k,t)=\case{\mathrm{d}}{\mathrm{d}t} \mathfrak{a}(k,t)$. The first term is represented by the left graph in Figure\ \ref{firstOrder:graphs}.
Let us first explain the structure of the graph. Each graph consists of the following symbols: vertices, edges and time slices. The time direction points from bottom to top. The $n$-th order terms have $n$ vertices, and so the first order terms have only a single vertex. The vertex represents the interaction of particles. The edges are labeled by oriented momentum-variables $k_i$. If the earlier of the endpoints is a creation operator, the arrow points in the time direction, and if it is an annihilation operator, the arrow points opposite to the time direction. Then, by definition of $\mathcal{A}$, at every vertex there are two ingoing and two outgoing arrows. 

To reconstruct the correspondig integral from a given graph, one needs to iteratively add the following five operations for each vertex:
\begin{enumerate}
	\item An integration of a time variable $s$ from zero to the end of the time slice after the vertex. In Figure\ \ref{firstOrder:graphs} this amounts to using the time interal $\int_0^t \mathrm{d}s$.
	\item The integration over the momentum variables can be read of as follows: one needs to add $\int_{(\mathbb{T}^d)^3} \mathrm{d}k_{ijl}$ where  $k_i$, $k_j$ and $k_l$ label the three ``earlier'' edges.
	\item A product of four phase factors $\mathrm{e}^{\pm \mathrm{i} \omega(k_j) s}$, one for each arrow attached to the vertex, where $s$ denotes the time integration variable of the vertex. A negative sign is chosen if the arrow points in the time direction, and a positive sign if it points against the time direction. 
	\item A $\delta$-function ensuring the momentum conservation, in which a positive sign is used if the corresponding arrow points away from the vertex, and a negative sign if the arrow points towards the vertex. 
	\item A factor ``$\pm \mathrm{i}$'' with a positive sign if the single later edge points away from the vertex, and a negative sign if it points towards the vertex.
\end{enumerate}

Finally, an average $\langle \cdot \rangle$ over the initial state needs to be taken of the product of creation and annihilation operators at the bottom of the graph. Also for every $(\hat{a}(k_i)^* \cdot \hat{a}(k_j))$ represents an influence of $\hat{V}(k_i - k_j)$. By construction, if one starts to count the direction of the arrows from left to right in any of the time slices, they always start with an up-arrow and alternate from left to right in up-down combinations. This results in an alternating sequence of creation and annihilation operators at the bottom of the graph. The Wick-pairings ``$\sqcup$'' shown under the graph follow from averaging this alternating sequence over the initial quasi-free state.  The average has a particularly simple form for the alternating order of creation and annihilation operators: it can then be computed according to the Wick rule
\begin{equation}
\langle \hat{a}_{i_1}^* \hat{a}_{j_1} \cdots \hat{a}_{i_n}^* \hat{a}_{j_n} \rangle = \mathrm{det}[K(i_k,j_l)]_{1 \leq k,l \leq n},	
\end{equation}
where
\begin{equation}
	K(i_k,j_l) = \left\{ \begin{array}{c c l} \langle \hat{a}_{i_k}^* \hat{a}_{j_l} \rangle & \mathrm{, if} & k \leq l, \\
	- \langle \hat{a}_{j_l} \hat{a}_{i_k}^* \rangle & \mathrm{, if} & k > l. \end{array} \right.
\end{equation}
For instance, the expectation value $\langle \cdot \rangle$ over the initial state in the first term in (\ref{eq:full1storder}) can be expressed as
\begin{eqnarray} \label{wick:det}
\fl 
\langle \big( \hat{a}(k_2)^* \cdot \hat{a}(k_3) \big) \big(\hat{a}_\mathrm{f}(k_4)^* \odot \hat{a}_{\mathrm{g}}(k_5) \big) \rangle \nonumber \\
= \sum_{\sigma_1, \sigma, \tau \in \{ \uparrow, \downarrow \}} \overline{\mathrm{f}}_\sigma \mathrm{g}_\tau 
		\langle \hat{a}_{\sigma_1}(k_2)^* \hat{a}_{\sigma_1}(k_3) \hat{a}_{\sigma}(k_4)^* \hat{a}_{\tau}(k_5) \rangle \nonumber  \\
= \sum_{\sigma_1, \sigma, \tau \in \{ \uparrow, \downarrow \}} \overline{\mathrm{f}}_\sigma \mathrm{g}_\tau \, \mathrm{det} \left[ 
	\begin{array}{c c c} 
				\langle \hat{a}_{\sigma_1}(k_2)^* \hat{a}_{\sigma_1}(k_3) \rangle &
				\langle \hat{a}_{\sigma_1}(k_2)^* \hat{a}_{\tau}(k_5) \rangle \nonumber \\
				- \langle \hat{a}_{\sigma_1}(k_3) \hat{a}_{\sigma}(k_4)^* \rangle &
				\langle \hat{a}_{\sigma}(k_4)^* \hat{a}_{\tau}(k_5) \rangle
	\end{array} \right] \\
= \sum_{\sigma_1, \sigma, \tau \in \{ \uparrow, \downarrow \}} \overline{\mathrm{f}}_\sigma \mathrm{g}_\tau \big( \langle \hat{a}_{\sigma_1}(k_2)^* \hat{a}_{\sigma_1}(k_3) \rangle \langle \hat{a}_{\sigma}(k_4)^* \hat{a}_{\tau}(k_5) \rangle \nonumber \\  \quad 
+ \langle \hat{a}_{\sigma_1}(k_3) \hat{a}_{\sigma}(k_4)^* \rangle \langle \hat{a}_{\sigma_1}(k_2)^* \hat{a}_{\tau}(k_5) \rangle \big).
\end{eqnarray}
The two Wick pairings shown in figure~\ref{firstOrder:graphs} represent the two different pairings in equation (\ref{wick:det}).  Since for instance, $\langle \hat{a}_{\sigma_1}(k_3) \hat{a}_\sigma(k_4)^*\rangle = \delta(k_3-k_4) \tilde{W}(k_4)_{\sigma\sigma_1}$, the left diagram yields
\begin{eqnarray}
\fl
	\int_0^t \mathrm{d}s \, \langle \dot{\mathfrak{a}}^*_{\mathrm{f}}(k_1,s)^{(1)} \odot \mathfrak{a}_{\mathrm{g}}(k_5,s)^{(0)} \rangle = \nonumber \\ 
	\mathrm{i} t \, \delta(k_1 - k_5) \int_{\mathbb{T}^{d}} \mathrm{d} k_2 \big( \hat{V}(0) \, \mathrm{tr}[W_2] \langle \mathrm{f}, W_1 \mathrm{g} \rangle + \hat{V}(k_1-k_2) \langle \mathrm{f}, \tilde{W}_2 W_1 \mathrm{g} \rangle \big).
\end{eqnarray}
The contribution of the right diagram in figure\ \ref{firstOrder:graphs} can also be computed directly by taking an adjoint of the result above, yielding
\begin{eqnarray}
\fl \int_0^t \mathrm{d}s \, \langle \mathfrak{a}^*_{\mathrm{f}}(k_1,s)^{(0)} \odot \dot{\mathfrak{a}}_{\mathrm{g}}(k_5,s)^{(1)} \rangle = \nonumber \\ 
	- \mathrm{i} t \, \delta(k_1 - k_5) \int_{\mathbb{T}^{d}} \mathrm{d} k_2 \big( \hat{V}(0) \, \mathrm{tr}[W_2] \langle \mathrm{f}, W_1 \mathrm{g} \rangle 
		+ \hat{V}(k_1 - k_2) \langle \mathrm{f}, W_1 \tilde{W}_2 \mathrm{g} \rangle \big).	
\end{eqnarray}
Thus the first order term is given by
\begin{equation}
\fl W^{(1)}(k_1,t) = - \mathrm{i} t \, [ R[W]_1, W_1 ], \qquad
	R[W]_1 = \int_{\mathbb{T}^{d}} \mathrm{d} k \, \hat{V}(k_1 - k) \, W(k) \in \mathbb{C}_{2 \times 2}.
\end{equation}
All four diagrams in figure\ \ref{firstOrder:graphs} have an interaction with zero momentum transfer (for instance, using the top left pairing leads to $k_4 = k_1$). Such diagrams will also appear in the second order and we call them 
{\em zero momentum transfer diagrams\/}.
\section{Second order terms}
We next consider the second order term which we decompose into a sum of four terms, obtained by evaluating the time-derivative in the equality
\begin{equation} \label{W:2}
\fl \delta(k - \tilde{k}) \langle \mathrm{f}, \, W^{(2)}(k,t) \mathrm{g} \rangle = \int_0^t \mathrm{d}s
	\sum_{m=0}^2 \frac{\mathrm{d}}{\mathrm{d}s} 
	\langle \mathfrak{a}^*_\mathrm{f}(k,s)^{(m)} \odot \mathfrak{a}_\mathrm{g}(\tilde{k},s)^{(2-m)} \rangle.
\end{equation}
\subsection*{(1$\bf{'}$,1)-term:}
\begin{figure}
\centering
\includegraphics[width=0.7\columnwidth]{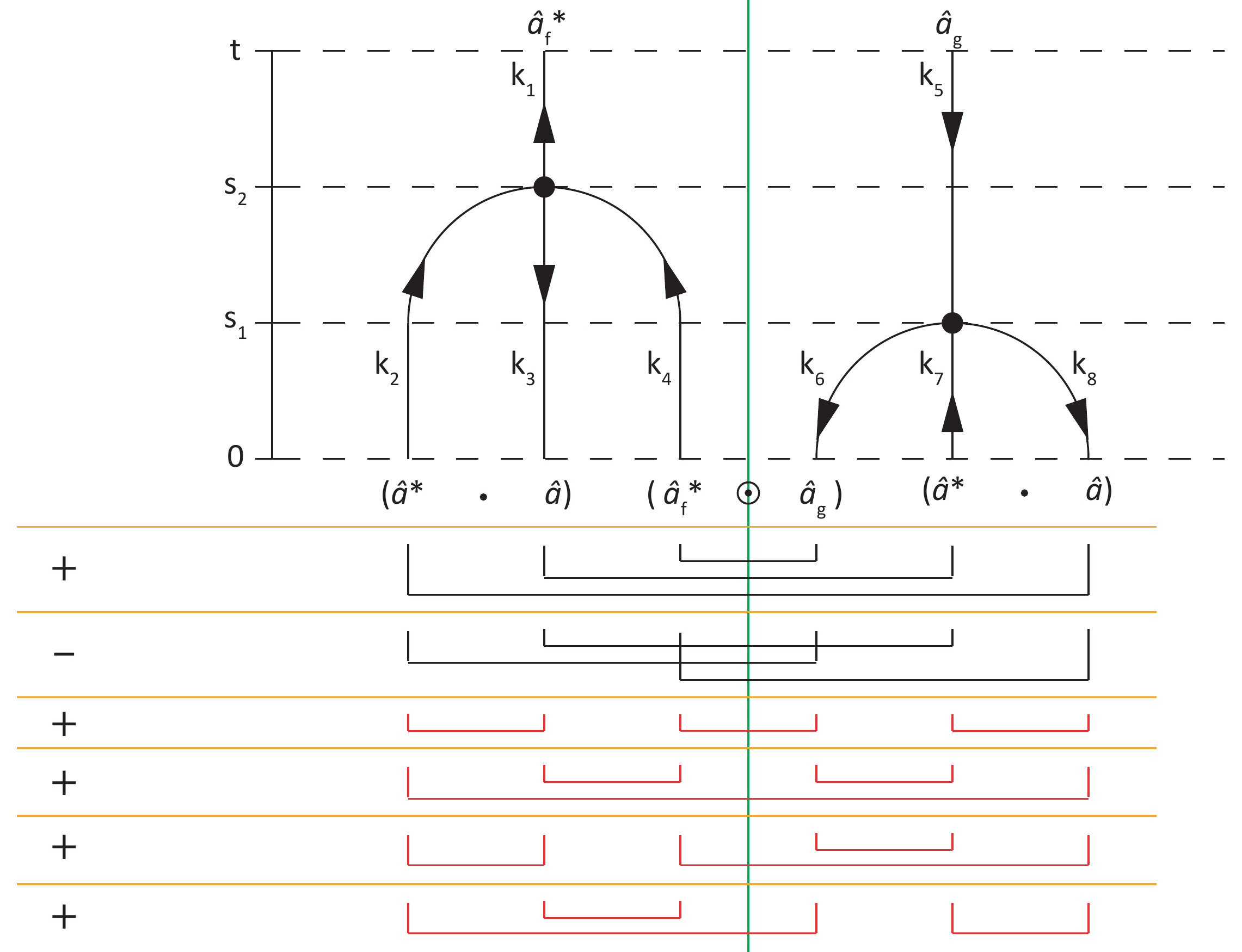}
\caption{The diagrams of the $(1',1)$-terms.} \label{secondOrder:graph:1}
\end{figure}

\noindent In the previous section we have already shown that
\begin{eqnarray}\label{eq:oneponeterm}
\fl \int_0^t \mathrm{d}s \, \langle \dot{\mathfrak{a}}^*_\mathrm{f}(k_1,s)^{(1)} \odot \mathfrak{a}_\mathrm{g}(k_5,s)^{(1)} \rangle = \nonumber \\
	\int_0^t \mathrm{d}s_2 \int_{0}^{s_2} \mathrm{d}s_1 \, \langle 
	\mathcal{A}_*[\mathrm{e}^{-\mathrm{i} \omega_{1234} s_2}, \mathfrak{a}^*, \mathfrak{a}, \mathfrak{a}^*_\mathrm{f}](k_1) \odot 
	\mathcal{A}[\mathrm{e}^{\mathrm{i} \omega_{5678} s_1}, \mathfrak{a}_\mathrm{g}, \mathfrak{a}^*, \mathfrak{a}](k_5) \rangle
\end{eqnarray}
which can be represented by the Feynman diagram of Figure\ \ref{secondOrder:graph:1}.
%
In order to evaluate the diagram we start with
\begin{eqnarray}
\fl \langle \big( \hat{a}(k_2)^* \cdot \hat{a}(k_3) \big) \big( \hat{a}_\mathrm{f}(k_4)^* \odot \hat{a}_\mathrm{g}(k_6) \big) 
	\big( \hat{a}(k_7)^* \cdot \hat{a}(k_8) \big) \rangle \nonumber \\
	= \sum_{\sigma,\tau,\mu_1,\mu_2} \overline{\mathrm{f}}_\sigma \mathrm{g}_\tau 
	\langle \hat{a}_{\mu_1}(k_2)^* \hat{a}_{\mu_1}(k_3) \hat{a}_\sigma(k_4)^* \hat{a}_\tau(k_6) 
	\hat{a}_{\mu_2}(k_7)^* \hat{a}_{\mu_2}(k_8) \rangle. 
\end{eqnarray}
Using
\begin{eqnarray}
\fl \langle \hat{a}_{s_1}(i_1)^* \hat{a}_{r_1}(j_1) \hat{a}_{s_2}(i_2)^* 
	  \hat{a}_{r_2}(j_2) \hat{a}_{s_3}(i_3)^* \hat{a}_{r_3}(j_3) \rangle  \nonumber \\
 =  \mathrm{det} \left[ \begin{array}{c c c} 
		\langle \hat{a}_{s_1}(i_1)^* \hat{a}_{r_1}(j_1) \rangle &
		\langle \hat{a}_{s_1}(i_1)^* \hat{a}_{r_2}(j_2) \rangle &
		\langle \hat{a}_{s_1}(i_1)^* \hat{a}_{r_3}(j_3) \rangle \\
		- \langle \hat{a}_{r_1}(j_1) \hat{a}_{s_2}(i_2)^* \rangle &
		\langle \hat{a}_{s_2}(i_2)^* \hat{a}_{r_2}(j_2) \rangle &
		\langle \hat{a}_{s_2}(i_2)^* \hat{a}_{r_3}(j_3) \rangle \\
		- \langle \hat{a}_{r_1}(j_1) \hat{a}_{s_3}(i_3)^* \rangle &
		- \langle \hat{a}_{r_2}(j_2) \hat{a}_{s_3}(i_3)^* \rangle &
		\langle \hat{a}_{s_3}(i_3)^* \hat{a}_{r_3}(j_3) \rangle
	 \end{array} \right], 
\end{eqnarray}
one arrives at
\begin{eqnarray}
\fl \langle \big( \hat{a}(k_2)^* \cdot \hat{a}(k_3) \big) 
	\big( \hat{a}_\mathrm{f}(k_4)^* \odot \hat{a}_\mathrm{g}(k_6) \big) \big( \hat{a}(k_7)^* \cdot \hat{a}(k_8) \big) \rangle \nonumber \\
=\delta(k_3 - k_7) \delta(k_4 - k_6) \delta(k_2 - k_8) \, \langle \mathrm{f}, W_4 \mathrm{tr}[\tilde{W}_3 W_2] \mathrm{g} \rangle \nonumber \\
\quad
- \delta(k_2 - k_6) \delta(k_4 - k_8) \delta(k_3 - k_7) \, \langle \mathrm{f}, W_4 \tilde{W}_3 W_2 \mathrm{g} \rangle 
\nonumber \\ \quad 
+ \delta(k_2 - k_3) \delta(k_4 - k_6) \delta(k_7 - k_8) \, \langle \mathrm{f}, W_4 \mathrm{tr}[W_2] \mathrm{tr}[W_7] \mathrm{g} \rangle \nonumber \\ \quad
+ \delta(k_2 - k_8) \delta(k_3 - k_4) \delta(k_6 - k_7) \, \langle \mathrm{f}, \tilde{W}_4 W_2 \tilde{W}_6 \mathrm{g} \rangle \nonumber \\ \quad
+ \delta(k_6 - k_7) \delta(k_4 - k_8) \delta(k_2 - k_3) \, \langle \mathrm{f}, W_4 \tilde{W}_6 \mathrm{tr}[W_2] \mathrm{g} \rangle \nonumber \\ \quad
+ \delta(k_7 - k_8) \delta(k_3 - k_4) \delta(k_2 - k_6) \, \langle \mathrm{f}, \tilde{W}_3 W_2 \mathrm{tr}[W_7] \mathrm{g} \rangle. \label{exp:1}
\end{eqnarray}
Using this formula in (\ref{eq:oneponeterm}) yields the following expression for the $(1',1)$-term,
\begin{eqnarray}
\fl \int_0^t \mathrm{d}s \, \langle \dot{\mathfrak{a}}^*_\mathrm{f}(k_1,s)^{(1)} \odot \mathfrak{a}_\mathrm{g}(k_5,s)^{(1)} \rangle 
	= \delta(k_1 - k_5) \, \frac{1}{2} t^2 \, \langle \mathrm{f}, \, \mathcal{Z}[W]_1^{(1' 1)} \mathrm{g} \rangle \nonumber \\ 
\fl \qquad + \delta(k_1 - k_5) \int_0^t \mathrm{d}s_2 \int_0^{s_2} \mathrm{d}s_1 \, 
		\int_{(\mathbb{T}^{d})^3} \mathrm{d} k_{234} \, \delta(\underline{k})
		\mathrm{e}^{-\mathrm{i} \omega_{1234} (s_2-s_1)}
		\langle \mathrm{f}, \, \mathcal{D}[W]^*_{234} \mathrm{g} \rangle. 
\end{eqnarray}
Here
\begin{equation}
\fl \mathcal{D}[W]^*_{234} = \hat{V}(k_2 - k_3)^2 \, W_4 \mathrm{tr}[\tilde{W}_3 W_2] - \hat{V}(k_2 - k_3) \hat{V}(k_3 - k_4) \, W_4 \tilde{W}_3 W_2,  	
\end{equation}
and it results from the first two terms in equation (\ref{exp:1}). The remaining four terms all lead to a diagram with a zero momentum transfer and summing up their contribution yields
\begin{equation}
\fl \mathcal{Z}[W]_1^{(1' 1)} = \hat{V}(0) \{ W_1, \, R[\tilde{W}]_1 \} \, \mathrm{tr}[R] + R[\tilde{W}]_1 \, W_1 \, R[\tilde{W}]_1 + \hat{V}(0)^2 \, W_1 \, \mathrm{tr}[R] \mathrm{tr}[R].	
\end{equation} 
\subsection*{(1,1$\bf{'}$)-term:}
\noindent A similar discussion applies to
\begin{eqnarray}
\fl \int_0^t \mathrm{d}s \, \langle \mathfrak{a}^*_\mathrm{f}(k_1,s)^{(1)} \odot \dot{\mathfrak{a}}_\mathrm{g}(k_5,s)^{(1)} \rangle  \nonumber \\
\fl \qquad = \int_0^t \mathrm{d}s_2 \int_0^{s_2} \mathrm{d}s_1 \,  
	\langle \mathcal{A}_*[\mathrm{e}^{-\mathrm{i} \omega_{1234} s_1}, \mathfrak{a}^*, \mathfrak{a}, \mathfrak{a}^*_\mathrm{f}](k_1) \odot 
	\mathcal{A}[\mathrm{e}^{\mathrm{i} \omega_{5678} s_2}, \mathfrak{a}_\mathrm{g}, \mathfrak{a}^*, \mathfrak{a}](k_5) \rangle,
\end{eqnarray}
which can also be computed by taking the adjoint of the $(1', 1)$-term.  This shows that 
\begin{eqnarray}
\fl \int_0^t \mathrm{d}s \, \langle \mathfrak{a}^*_\mathrm{f}(k_1,s)^{(1)} \odot \dot{\mathfrak{a}}_\mathrm{g}(k_5,s)^{(1)} \rangle
	= \delta(k_1 - k_5) \, \frac{1}{2} t^2 \, \langle \mathrm{f}, \, \mathcal{Z}[W]_1^{(1 1')} \mathrm{g} \rangle \nonumber \\
\fl \qquad + \delta(k_1 - k_5) \int_0^t \mathrm{d}s_2 \int_0^{s_2} \mathrm{d}s_1 \int_{(\mathbb{T}^{d})^3} \mathrm{d} k_{234} \, 
	\delta(\underline{k}) \, \mathrm{e}^{\mathrm{i} \omega_{1234} (s_2-s_1)} \langle \mathrm{f}, \, \mathcal{D}[W]_{234} \mathrm{g} \rangle,
\end{eqnarray}
where $\mathcal{Z}[W]_1^{(1 1')}= (\mathcal{Z}[W]_1^{(1' 1)})^* = \mathcal{Z}[W]_1^{(1' 1)} $.
\subsection*{(2,0)-term:}
\begin{figure}
	\begin{center}
	\begin{minipage}[t]{0.4 \textwidth} \centering
		\includegraphics[width=1\columnwidth]{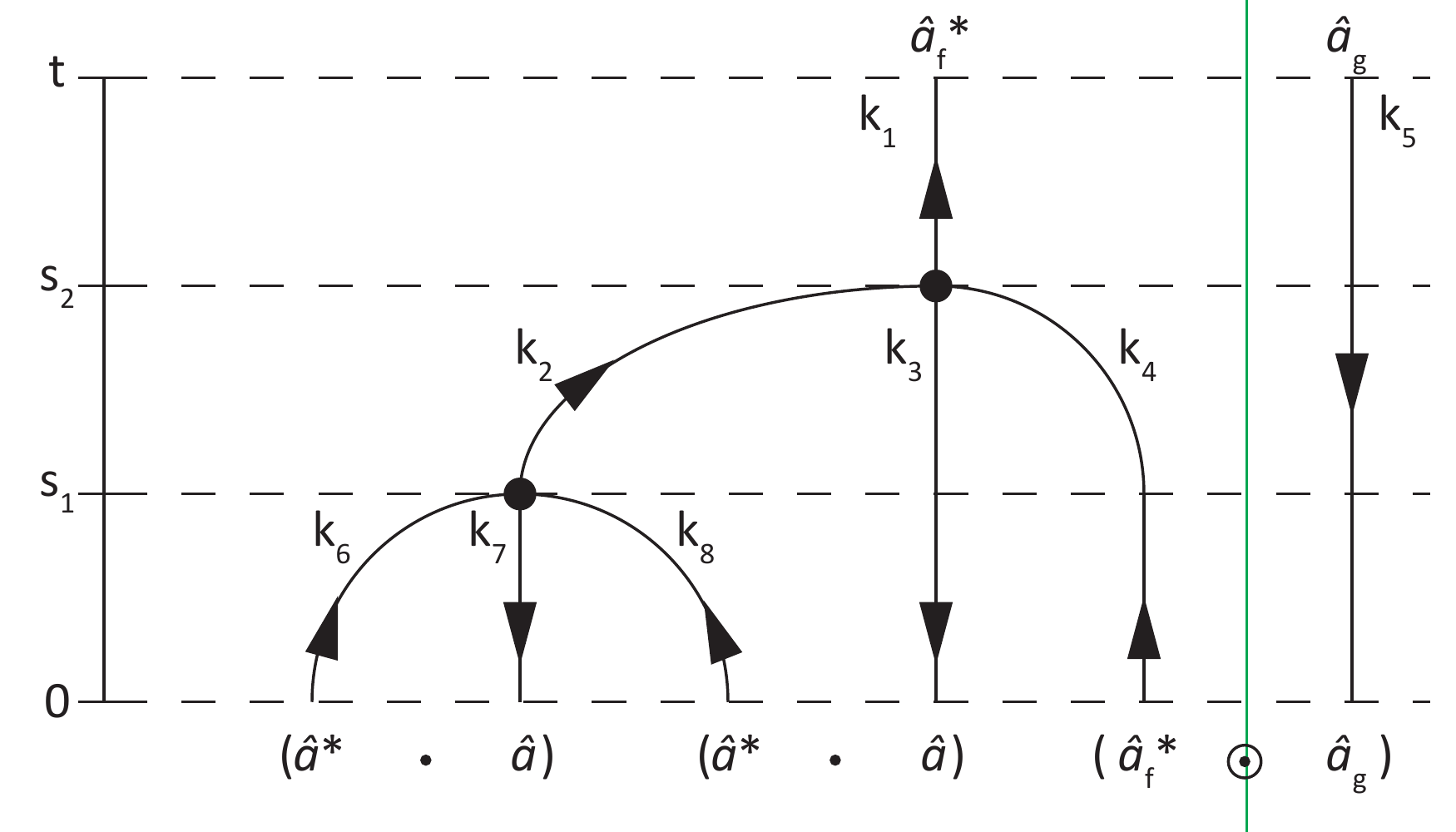} \\ (a)
	\end{minipage}
	\quad                    
	\begin{minipage}[t]{0.4 \textwidth} \centering
		\includegraphics[width=1\columnwidth]{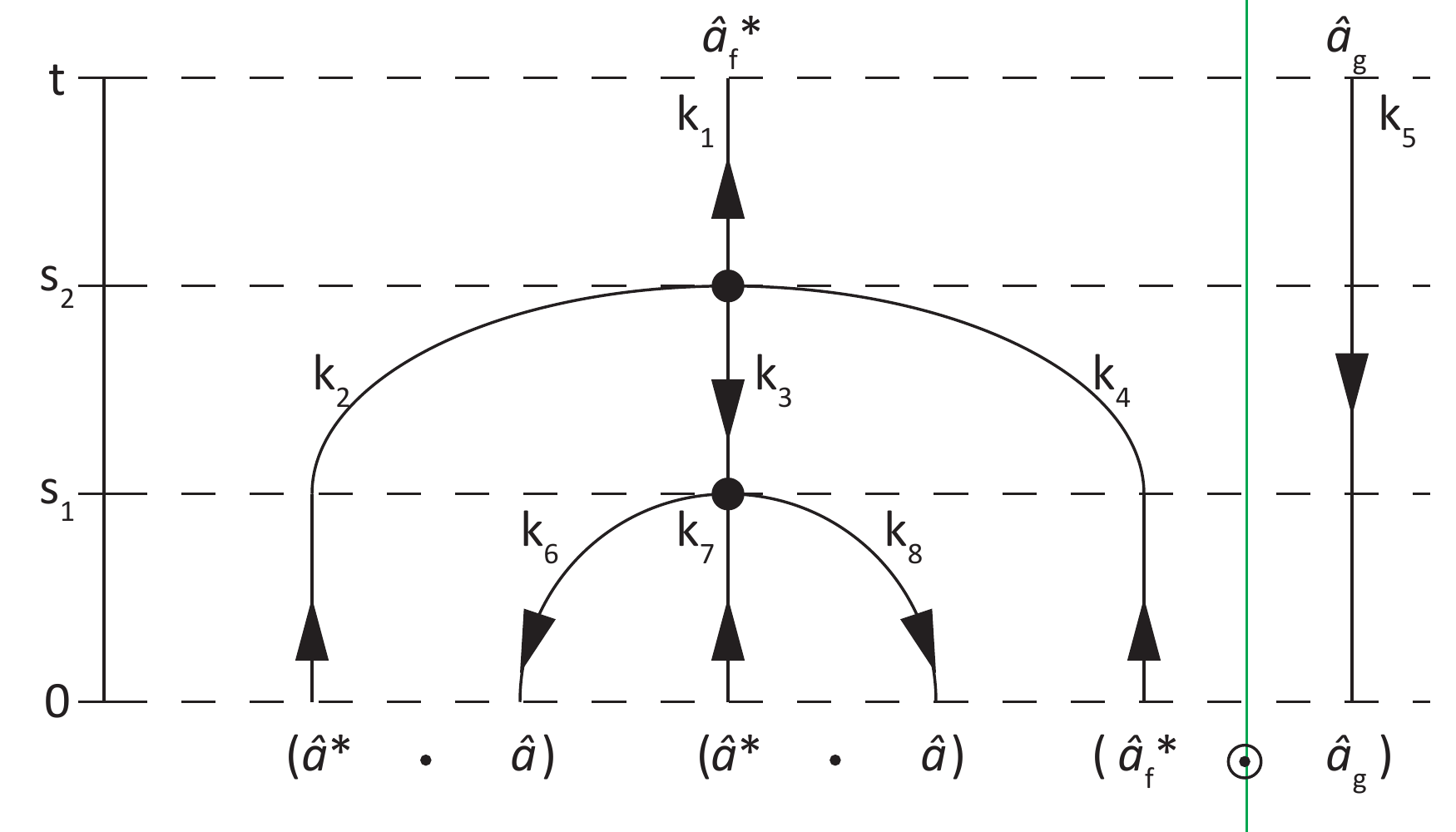} \\ (b)
	\end{minipage}
	\begin{minipage}[t]{0.4 \textwidth} \centering
		\includegraphics[width=1\columnwidth]{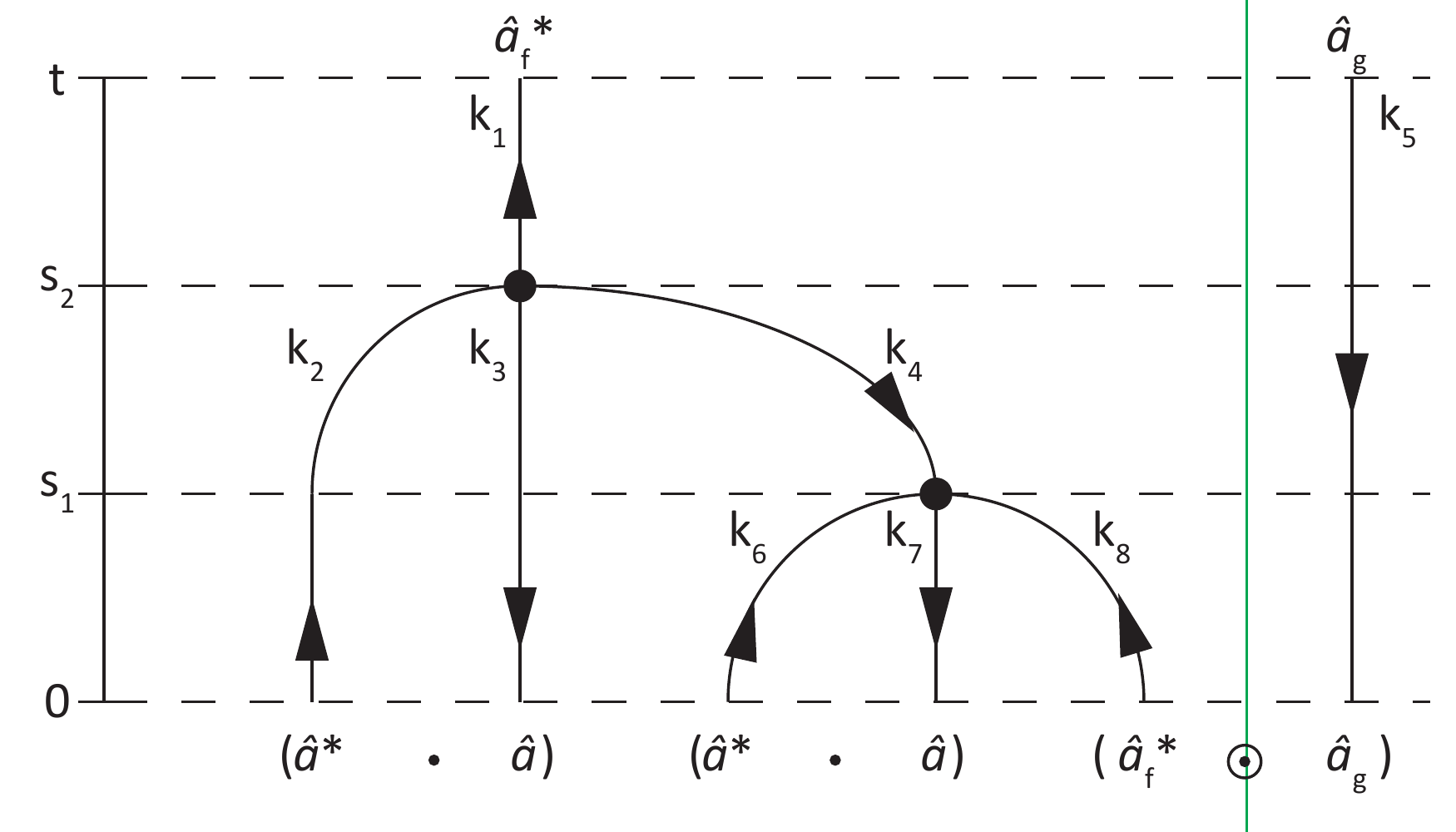} \\ (c)
	\end{minipage}
	\end{center}
	\caption{Graphs related to the three terms in (\ref{exp:20:3}): (a) first term, (b) second term, (c) third term.\label{secondOrder:graph21}}
\end{figure}

\noindent The $(2,0)$-term is given by the following expression
\begin{eqnarray}
\fl \int_0^t \mathrm{d}s \, \langle \dot{\mathfrak{a}}^*_\mathrm{f}(k_1,s)^{(2)} \odot \mathfrak{a}_\mathrm{g}(k_5,s)^{(0)} \rangle = \nonumber \\
\fl \qquad - \int_0^t \mathrm{d}s_2 \int_0^{s_2} \mathrm{d}s_1 \, 
	\langle \mathcal{A}_*[\mathrm{e}^{-\mathrm{i} \omega_{1234} s_2}, \mathcal{A}_*[\mathrm{e}^{-\mathrm{i} \omega_{2678} s_1},
		\mathfrak{a}^*, \mathfrak{a}, \mathfrak{a}^*], \mathfrak{a}, \mathfrak{a}^*_\mathrm{f}](k_1) \odot \mathfrak{a}_\mathrm{g}(k_5) \rangle
			\label{exp:20:1} \nonumber \\
\fl \qquad + \int_0^t \mathrm{d}s_2 \int_0^{s_2} \mathrm{d}s_1 \, 
	\langle \mathcal{A}_*[\mathrm{e}^{-\mathrm{i} \omega_{1234} s_2}, \mathfrak{a}^*, \mathcal{A}[\mathrm{e}^{\mathrm{i} \omega_{3678} s_1},
		\mathfrak{a}, \mathfrak{a}^*, \mathfrak{a}], \mathfrak{a}^*_\mathrm{f}](k_1) \odot \mathfrak{a}_\mathrm{g}(k_5) \rangle 
			\label{exp:20:2} \nonumber \\
\fl \qquad - \int_0^t \mathrm{d}s_2 \int_0^{s_2} \mathrm{d}s_1 \, 
	\langle \mathcal{A}_*[\mathrm{e}^{-\mathrm{i} \omega_{1234} s_2}, \mathfrak{a}^*, \mathfrak{a}, 
		\mathcal{A}_*[ \mathrm{e}^{-\mathrm{i} \omega_{4678} s_1}, \mathfrak{a}^*, \mathfrak{a}, \mathfrak{a}^*_\mathrm{f}]](k_1) 
		\odot \mathfrak{a}_\mathrm{g}(k_5) \rangle. \label{exp:20:3}
\end{eqnarray}
The associated graphs are shown in Figure\ \ref{secondOrder:graph21}.
To evaluate the contribution of the pairings to the first term in equation (\ref{exp:20:1}) we use
\begin{eqnarray}
\fl \langle \big( \hat{a}(k_6)^* \cdot \hat{a}(k_7) \big) \big( \hat{a}(k_8)^* \cdot \hat{a}(k_3) \big) 
		\big( \hat{a}_\mathrm{f}(k_4)^* \odot \hat{a}_\mathrm{g}(k_5) \big) \rangle \nonumber \\
  = \delta(k_7 - k_4) \delta(k_8 - k_3) \delta(k_6 - k_5) \, \langle \mathrm{f}, \, \tilde{W}_4 W_5 \mathrm{tr}[W_3] \mathrm{g} \rangle \nonumber \\ 
  \quad - \delta(k_6 - k_3) \delta(k_8 - k_5) \delta(k_7 - k_4) \, \langle \mathrm{f}, \, \tilde{W}_4 W_3 W_5 \mathrm{g} \rangle \nonumber \\
  \quad + \textrm{ zero momentum transfer diagrams}  .
\end{eqnarray}
The contribution to the second term in equation (\ref{exp:20:2}) can be computed using
\begin{eqnarray}
\fl \langle \big( \hat{a}(k_2)^* \cdot \hat{a}(k_6) \big) \big( \hat{a}(k_7)^* \cdot \hat{a}(k_8) \big) 
	\big( \hat{a}_\mathrm{f}(k_4)^* \odot \hat{a}_\mathrm{g}(k_5) \big) \rangle \nonumber \\
  = \delta(k_8 - k_4) \delta(k_7 - k_5) \delta(k_6 - k_2) \, \langle \mathrm{f}, \, \tilde{W}_4 W_5 \mathrm{tr}[W_2] \mathrm{g} \rangle \nonumber \\ 
  \quad - \delta(k_2 - k_8) \delta(k_7 - k_5) \delta(k_6 - k_4) \, \langle \mathrm{f}, \, \tilde{W}_4 W_2 W_5 \mathrm{g} \rangle \nonumber \\
  \quad + \textrm{ zero momentum transfer diagrams} ,
\end{eqnarray}
and the contribution to the third term in equation (\ref{exp:20:3}) by
\begin{eqnarray}
\fl \langle \big( \hat{a}(k_2)^* \cdot \hat{a}(k_3) \big) \big( \hat{a}(k_6)^* \cdot \hat{a}(k_7) \big) 
\big( \hat{a}_\mathrm{f}(k_8)^* \odot \hat{a}_\mathrm{g}(k_5) \big) \rangle \nonumber \\
  = \delta(k_8 - k_5) \delta(k_3 - k_6) \delta(k_2 - k_7) \, \langle \mathrm{f}, \, W_5 \mathrm{tr}[\tilde{W}_3 W_2 ] \mathrm{g} \rangle \nonumber \\ 
  \quad - \delta(k_2 - k_7) \delta(k_6 - k_5) \delta(k_3 - k_8) \, \langle \mathrm{f}, \, \tilde{W}_3 W_2 W_5 \mathrm{g} \rangle \nonumber \\
  \quad + \textrm{ zero momentum transfer diagrams}. 
\end{eqnarray}
With the definitions
\begin{eqnarray}
\fl \mathcal{B}[W]^*_{1234} = \hat{V}(k_2 - k_3) \hat{V}(k_3-k_4) 
	\Big( \tilde{W}_4 W_3 W_{1} + \tilde{W}_3 W_2 W_{1} - \tilde{W}_4 W_2 W_{1} \Big) \nonumber \\ 
	+ \hat{V}(k_2 - k_3)^2 \Big( \tilde{W}_4 W_{1} \mathrm{tr}[W_2] - \tilde{W}_4 W_{1} \mathrm{tr}[W_3] - W_{1} \mathrm{tr}[\tilde{W}_3 W_2] \Big)
\end{eqnarray}
and
\begin{eqnarray}
\fl \mathcal{Z}[W]_1^{(2 0)} = 
	- \hat{V}(0)^2 \, W_1 \, \mathrm{tr}[R] \mathrm{tr}[R] 
	- R[\tilde{W}]_1 \, R[\tilde{W}]_1 \, W_1
	- \hat{V}(0) \, R[\tilde{W}]_1 \, W_1 \, \mathrm{tr}[R] \nonumber \\
	- \hat{V}(0) \, R[\tilde{W}]_1 \, W_1 \, \mathrm{tr}[R]
\end{eqnarray}
we obtain
\begin{eqnarray}\label{eq:twozeroresult}
\fl \int_0^t \mathrm{d}s \, \langle \dot{\mathfrak{a}}^*_\mathrm{f}(k_1,t)^{(2)} \odot \mathfrak{a}_\mathrm{g}(k_5,t)^{(0)} \rangle
	= \delta(k_1 - k_5) \frac{1}{2} t^2 \, \langle \mathrm{f}, \, \mathcal{Z}[W]_1^{(2 0)} \mathrm{g} \rangle \nonumber \\
\fl \qquad + \delta(k_1 - k_5) \int_0^t \mathrm{d}s_1 \int_0^{s_1} \mathrm{d}s_2 \int_{(\mathbb{T}^{d})^3} \mathrm{d} k_{234} \, \delta(\underline{k}) \, 
		\mathrm{e}^{-\mathrm{i} \omega_{1234} (s_2-s_1)} \langle \mathrm{f}, \, \mathcal{B}[W]^*_{1234} \, \mathrm{g} \rangle.
\end{eqnarray}
\subsection*{(0,2)-term:}
\noindent The $(0,2)$-term is given by the following expression
\begin{eqnarray}
\fl \int_0^t \mathrm{d}s \, \langle \mathfrak{a}^*_\mathrm{f}(k_1,s)^{(0)} \odot \dot{\mathfrak{a}}_\mathrm{g}(k_5,s)^{(2)} \rangle = \nonumber \\
\fl \qquad - \int_0^t \mathrm{d}s_2 \int_0^{s_2} \mathrm{d}s_1 \, \langle \mathfrak{a}^*_\mathrm{f}(k_1) \odot 
	\mathcal{A}[\mathrm{e}^{\mathrm{i} \omega_{5234} s_2},\mathcal{A}[\mathrm{e}^{\mathrm{i} \omega_{2678} s_1}, \mathfrak{a}_\mathrm{g},
	\mathfrak{a}^*, \mathfrak{a}],\mathfrak{a}^*,\mathfrak{a}](k_5) \rangle \label{exp:02:1} \nonumber \\
\fl \qquad + \int_0^t \mathrm{d}s_2 \int_0^{s_2} \mathrm{d}s_1 \, \langle \mathfrak{a}^*_\mathrm{f}(k_1) \odot 
	\mathcal{A}[\mathrm{e}^{\mathrm{i} \omega_{5234} s_2},\mathfrak{a}_\mathrm{g},
	\mathcal{A}_*[\mathrm{e}^{-\mathrm{i} \omega_{3678} s_1},\mathfrak{a}^*,\mathfrak{a},\mathfrak{a}^*],\mathfrak{a}](k_5) \rangle
		\label{exp:02:2} \nonumber \\
\fl \qquad - \int_0^t \mathrm{d}s_2 \int_0^{s_2} \mathrm{d}s_1 \, \langle \mathfrak{a}^*_\mathrm{f}(k_1) \odot 
	\mathcal{A}[\mathrm{e}^{\mathrm{i} \omega_{5234} s_2}, \mathfrak{a}_\mathrm{g}, \mathfrak{a}^*, 
	\mathcal{A}[\mathrm{e}^{\mathrm{i} \omega_{4678} s_1},\mathfrak{a},\mathfrak{a}^*,\mathfrak{a}]](k_5) \rangle.
		\label{exp:02:3}
\end{eqnarray}
The correponding graphs are computed  similarly to the $(2,0)$-term, and can also be obtained by reflecting each of the three graphs in figure\ \ref{secondOrder:graph21} at the vertical green (grey) line through $\odot$ and interchanging $\mathrm{f}$ and $\mathrm{g}$. By defining
\begin{eqnarray}
\fl \mathcal{B}[W]_{1234} = 
	\hat{V}(k_2 - k_3) \hat{V}(k_3 - k_4) \Big( W_1 W_4 \tilde{W}_3 + W_1 W_3 \tilde{W}_2 - W_1 W_4 \tilde{W}_2 \Big) \nonumber \\
	+ \hat{V}(k_3 - k_4)^2 \Big( W_1 \tilde{W}_2 \mathrm{tr}[W_4] 
	- W_1 \mathrm{tr}[\tilde{W}_3 W_4]  
	- W_1 \tilde{W}_2 \mathrm{tr}[W_3] \Big)	
\end{eqnarray}
one can show that the value of the integral in (\ref{eq:twozeroresult}) does not change if $\mathcal{B}[W]^*_{1234}$ is replaced there by $( \mathcal{B}[W]_{1234})^*$.  The contribution of the zero momentum transfer graphs is given by
\begin{eqnarray}
\fl \mathcal{Z}[W]_1^{(0 2)} = - \hat{V}(0)^2 \, W_1 \, \mathrm{tr}[R] \mathrm{tr}[R] - W_1 \, R[\tilde{W}]_1 \, R[\tilde{W}]_1
	- \hat{V}(0) \, W_1 \, R[\tilde{W}]_1 \, \mathrm{tr}[R] \nonumber \\
	- \hat{V}(0) \, W_1 \, R[\tilde{W}]_1 \, \mathrm{tr}[R] .
\end{eqnarray}
It holds that $(\mathcal{Z}[W]_1^{(20)})^* = \mathcal{Z}[W]_1^{(0 2)}$ and we finally obtain
\begin{eqnarray}
\fl \int_0^t \mathrm{d}s \, \langle \mathfrak{a}^*_\mathrm{f}(k_1,s)^{(0)} \odot \dot{\mathfrak{a}}_\mathrm{g}(k_5,s)^{(2)} \rangle
	= \delta(k_1 - k_5) \, \frac{1}{2} t^2 \, \langle \mathrm{f}, \, \mathcal{Z}[W]^{(0 2)} \mathrm{g} \rangle \nonumber \\
\fl \qquad + \delta(k_1 - k_5) \int_0^t \mathrm{d}s_2 \int_0^{s_2} \mathrm{d}s_1  \int_{(\mathbb{T}^{d})^3} \mathrm{d}^3 k_{234} \, 
		\delta(\underline{k}) \,
		\mathrm{e}^{\mathrm{i} \omega_{1234} (s_2-s_1)} \langle \mathrm{f}, \, \mathcal{B}[W]_{1234} \, \mathrm{g} \rangle \, .
\end{eqnarray}
\section{The limit $\lambda \rightarrow 0$, $t = \mathcal{O}(\lambda^{-2})$} \label{section3}
Before we consider the limit $\lambda \rightarrow 0$ we summarize all second order diagrams.
\noindent Defining
\begin{eqnarray}
\mathcal{A}[W]_\mathrm{1234} = \mathcal{D}[W]_\mathrm{234} + \mathcal{B}[W]_\mathrm{1234},\\
\mathcal{A}[W]^*_\mathrm{1234} = \mathcal{D}[W]^*_\mathrm{234} + \mathcal{B}[W]^*_\mathrm{1234},
\end{eqnarray}
and using the identity
\begin{equation}
\fl - [R[W]_1, \, [R[W]_1, \, W_1]] = \mathcal{Z}[W]_1^{(1' 1)} + \mathcal{Z}[W]_1^{(1 1')} 
	+ \mathcal{Z}[W]_1^{(2 0)} + \mathcal{Z}[W]_1^{(0 2)},
\end{equation}
we thus find that 
\begin{eqnarray}
\fl \int_0^t \mathrm{d}s \, \frac{\mathrm{d}}{\mathrm{d}s} 
	\sum_{m=0}^2 \langle \mathfrak{a}^*_\mathrm{f}(k_1,s)^{(m)} \odot \mathfrak{a}_\mathrm{g}(k_5,s)^{(2-m)} \rangle 
\nonumber \\ 
\fl \qquad = - \delta(k_1 - k_5) \, \frac{1}{2} t^2 \, \langle \mathrm{f}, \, [R[W]_1, \, [R[W]_1, \, W_1]] \mathrm{g} \rangle \nonumber \\ 
\fl \qquad\quad + \delta(k_1 - k_5) \int_0^t \mathrm{d}s_1 \int_0^{s_1} \mathrm{d}s_2 \int_{(\mathbb{T}^{d})^3} \mathrm{d} k_{234} \,
	\delta(\underline{k}) \, \mathrm{e}^{\mathrm{i} \omega_{1234} (s_2-s_1)} 
	\langle \mathrm{f}, \, \mathcal{A}[W]_{1234} \, \mathrm{g} \rangle \nonumber \\ 
\fl \qquad\quad + \delta(k_1 - k_5) \int_0^t \mathrm{d}s_1 \int_0^{s_1} \mathrm{d}s_2 \int_{(\mathbb{T}^{d})^3} \mathrm{d} k_{234} \, 
	\delta(\underline{k}) \, \mathrm{e}^{-\mathrm{i} \omega_{1234} (s_2-s_1)}
	\langle \mathrm{f}, \, \mathcal{A}[W]^*_{1234} \, \mathrm{g} \rangle .
\end{eqnarray}
Hence the second order term $W^{(2)}$ is given by
\begin{eqnarray}
W^{(2)}(k_1,t) = W^{(2)}_\mathrm{z}(k_1,t) + W^{(2)}_\mathrm{c}(k_1,t)
\end{eqnarray}
where
\begin{eqnarray}\label{eq:sumZ2}
W^{(2)}_\mathrm{z}(k_1,t)= - \frac{1}{2} t^2 \, [R[W]_1, \, [R[W]_1, \, W_1]], 
\end{eqnarray}
and
 \begin{eqnarray}\label{eq:W2cdef}
\fl W^{(2)}_\mathrm{c}(k_1,t) \nonumber \\
\fl \quad = \int_0^t \mathrm{d}s_1 \int_0^{s_1} \mathrm{d}s_2 \int_{(\mathbb{T}^{d})^3} \mathrm{d} k_{234} \, \delta(\underline{k})
	  \big( \mathrm{e}^{\mathrm{i} \omega_{1234} (s_1-s_2)} \mathcal{A}[W]_{1234}
	+ \mathrm{e}^{-\mathrm{i} \omega_{1234} (s_1-s_2)} \mathcal{A}[W]^*_{1234} \big).
\end{eqnarray}
The collision operator is determined by taking at second order the limit $\lambda \rightarrow 0$ and simultaneous long times $\lambda^{-2} t$ with $t$ of order $1$. More explicitly,
\begin{equation}
t \, \mathcal{C}[W^{(0)}](k) = \lim_{\lambda \rightarrow 0} \lambda^2 \, W^{(2)}_\mathrm{c}(k, \lambda^{-2} t),
\end{equation}
where $W^{(2)}_\mathrm{c}$ is defined in (\ref{eq:W2cdef}).
To evaluate the limit, we make use of
\begin{eqnarray}
\fl \lim_{\lambda \rightarrow 0} \lambda^2 \int_0^{\lambda^{-2} t} \mathrm{d}s_1 \int_0^{s_1} \mathrm{d}s_2 \, 
	\mathrm{e}^{\pm \mathrm{i} \omega_{1234} (s_1-s_2)} = t \int_0^\infty \mathrm{d}s \, 
	\mathrm{e}^{\pm \mathrm{i} \omega_{1234} s}  
	\nonumber \\
	= t \, \Big( \pm \mathrm{i} \, \mathcal{P}\left(\frac{1}{\omega_{1234}} \right) + \pi \, \delta(\omega_{1234}) \Big)  	
\end{eqnarray}
%
%
%
where $\mathcal{P}$ denotes the principal value integral, as defined in Section \ref{sec:intro}. This yields
\begin{eqnarray} \label{result}
\fl \lim_{\lambda \rightarrow 0} \lambda^2 \, W^{(2)}_\mathrm{c}(k,\lambda^{-2} \, t)  
 	= t \, \pi \int_{(\mathbb{T}^{d})^3} \mathrm{d} k_{234} \, 
		\delta(\underline{k}) \, \delta(\omega_{1234}) 
		\langle \mathrm{f}, \, (\mathcal{A}[W]_{1234} + \mathcal{A}[W]^*_{1234}) \mathrm{g} \rangle \nonumber \\ 
	\quad + t \, \mathrm{i} \int_{(\mathbb{T}^{d})^3} \mathrm{d} k_{234} \, \delta(\underline{k})
	 	\mathcal{P}\left(\frac{1}{\omega_{1234}} \right) 
		\langle \mathrm{f}, \, (\mathcal{A}[W]_{1234} - \mathcal{A}[W]^*_{1234}) \mathrm{g} \rangle .
\end{eqnarray}
This agrees with the result stated in the introduction.

We note that in case $W_{\sigma \tau}(k) = \delta_{\sigma \tau} W_\sigma(k)$ the term containing the principal part vanishes. 
The effective hamiltonian results from the twofold degeneracy of the unperturbed $H_0$.

\section{Conclusions}
The kinetic equation for the Hubbard model has two novel features. Firstly the Boltzmann $f$-function becomes in a natural way $2 \times 2$ matrix-valued. Furthermore, besides the conventional collision term there appears a conservative, Vlasov type term with an effective hamiltonian depending itself on $W(t)$.

Of course, the next goal would be to arrive at predictions based on kinetic theory. In \cite{FMS12a,FMS12b} we studied the one-dimensional model by numerically integrating the Boltzmann equation. In the spatially homogeneous case we find exponential convergence to the steady state and the related entropy increase. However, the family of stationary solutions depends on the precise form of the dispersion relation $\omega$ and is related to the integrability of the Hubbard chain. In the more mathematical investigation \cite{LMS12} we examine the role of the effective hamiltonian for $d\ge 3$. Because of the principal part, this term is in fact fairly singular and may induce rapid oscillations in the solution $W(k,t)$. In \cite{FMS12a,FMS12b,LMS12} we use an on-site interaction, which simplifies somewhat the structure of the kinetic equation, as explained in the Appendix.

From a theoretical perspective, one might wonder about the structure of the higher order diagrams. For example at order $\lambda^4$ one expects four types of diagrams: \vspace{2mm} \\
$(i)$ those vanishing in the kinetic limit, \vspace{2mm} \\
$(ii)$ non-vanishing and summing up to the term
\begin{equation} \label{iter:1}
\case{1}{2} \lambda^4 t^2 \mathcal{C}[\mathcal{C}[W]],	
\end{equation}
$(iii)$ zero momentum transfer diagrams summing up to
\begin{equation} \label{iter:2}
\case{1}{4!} \lambda^4 t^4 [R[W], [R[W], [R[W], [R[W],W]]]],	
\end{equation}
$(iv)$ the cross-terms of type $(ii)$ and $(iii)$. \vspace{2mm}

The structure of higher order terms has been investigated already by van Hove \cite{LH} for quantum fluids and by Prigogine \cite{PRIG} for classical anharmonic crystals. In fact, the diagrams become rather intricate, iterated oscillatory integrals and their asymptotic is difficult to handle, see \cite{HUG,HL,BCEP1,BCEP2} for more recent related work.

From the rigorous perspective, the best understood model seems to be the weakly nonlinear Schr\"odinger equation on a lattice with on-site interaction \cite{LS09}. In this work equilibrium time correlations are studied in the regime of small coupling and the limit equation is a version of the Boltzmann equation linearized at equilibrium. The $R$ matrix becomes just a number. Still it gives rise to rapid oscillations on the kinetic scale, and the analysis indicates that higher order terms will have a more complicated structure than anticipated in (\ref{iter:1}), (\ref{iter:2}). In particular, the higher order zero momentum transfer diagrams generate terms diverging on the kinetic time scale. Therefore, the strategy is to first subtract the rapid oscillations related to $R$ and to show that thereby, in a certain sense, the sum of all zero momentum transfer diagrams cancel each other with a sufficiently high precision. On a technical level, this separation is achieved by the pair truncation, as explained in Section 3 of \cite{LS09}. After pair truncation, the diagrams are indeed separated into $(i)$ and $(ii)$ and follow the anticipated pattern, \textit{i.e.}, at order $\lambda^{2n}$ the diagrams of $(ii)$ sum up to
\begin{equation}
\case{1}{n!} (\lambda^2 t)^n \mathcal{C}^{(n)}[W], \qquad n\textrm{-fold iteration}.	
\end{equation}
One might hope that a similar type of analysis can be achieved for the Hubbard model, but this will be a task for the future.

\ack

The research of M.~F\"urst was supported by the TU-M\"unchen as well as LMU-M\"unchen. He acknowledges the support of the University Observatory Munich once for science during the genesis of this work, especially H.~Lesch. He also acknowledges support by the DFG cluster of excellence `Origin and Structure of the Universe'.

The research of J.~Lukkarinen and P.~Mei was supported by the Academy of Finland and partially by the ERC Advanced Investigator Grant 227772.  We are also grateful to the Nordic Institute for Theoretical Physics (NORDITA), Stockholm, Sweden, to
the Erwin Schr\"odinger International Institute for Mathematical Physics (ESI), Vienna, Austria, 
and to the Banff International Research Station for Mathematical Innovation and Discovery (BIRS), Banff, Canada, for their hospitality during the workshops in which part of the research for the present work has been performed.  

\appendix

\section{The collision operator}\label{sec:appendix}

%
%
\noindent For the special case of an on-site interaction we show that the kinetic equation (\ref{eq:BoltzmannEquation:1}) agrees with the one in \cite{FMS12a,FMS12b,LMS12}. Since $\hat{V} = 1$, the starting point is
\begin{equation}
\mathcal{C}_\mathrm{c}[W(t)](k) = - \mathrm{i} \,[H_\mathrm{eff}(k,t), W(k,t)]
\end{equation}
with
\begin{eqnarray}
\fl H_{\mathrm{eff},1} = \int_{\mathbb{T}^3} \mathrm{d} k_{234} \, \delta(\underline{k}) \, \mathcal{P} \left(\case{1}{\underline{\omega}}\right) \nonumber \\
\times \big( W_2 W_4 - W_3 W_2 - W_2 W_3 - \mathrm{tr}[W_4] W_2 + \mathrm{tr}[W_3] W_2 + W_3 \big).
\end{eqnarray}
and
\begin{eqnarray}
\mathcal{C}_\mathrm{d}[W]_1 = \pi \int_{\mathbb{T}^3} \mathrm{d} k_{234} \, \delta(\underline{k}) \, \delta(\underline{\omega})\,
\big( \mathcal{A}[W]_{1234} + \mathcal{A}[W]_{1234}^* \big)
\end{eqnarray}
where
\begin{eqnarray}
\fl \mathcal{A}[W]_{1234} = - W_4 \tilde{W}_3 W_2 + W_4 \, \mathrm{tr}[ \tilde{W}_3 W_2 ]
- \big\{\tilde{W}_4 W_2 - \tilde{W}_4 W_3 - \tilde{W}_3 W_2 \nonumber \\
+ \tilde{W}_4 \, \mathrm{tr}[ W_3 ] - \tilde{W}_4 \, \mathrm{tr}[ W_2 ] + \mathrm{tr}[ W_2 \tilde{W}_3 ] \big\} W_1.
\end{eqnarray}
There are many possible representations of the collision operator. The goal here is to show that the form derived in Section 5, agrees with the simpler expressions given in equations (\ref{eq:Cc})--(\ref{eq:Heff}). 
To achieve the representation of the dissipative part given in (\ref{diss:jan}), we consider $\mathcal{A}[W]_{1234} + \mathcal{A}[W]^*_{1234}$ and add
%
%
the zero term
\begin{eqnarray}
\fl 
0 = W_1 W_2 \mathrm{tr}[W_3 W_4] - W_1 W_2 \mathrm{tr}[W_3 W_4] + \mathrm{tr}[W_4 W_3] W_2 W_1 
	- \mathrm{tr}[W_4 W_3] W_2 W_1 
\nonumber \\
+ W_1 W_2 W_3 W_4 - W_1 W_2 W_3 W_4 
		+ W_4 W_3 W_2 W_1 - W_4 W_3 W_2 W_1 .
\end{eqnarray}
Then we  use the symmetry $k_2 \leftrightarrow k_4$ to replace $\mathcal{A}[W]_{1234} + \mathcal{A}[W]^*_{1234}$ in the first integral in (\ref{result}) by 
\begin{eqnarray}
 \fl \tilde{W}_1 W_4 \mathcal{J}[\tilde{W}_3 W_2] 
	+ \mathcal{J}[W_2 \tilde{W}_3] W_4 \tilde{W}_1 
		- W_1 \tilde{W}_4 \mathcal{J}[W_3 \tilde{W}_2] - \mathcal{J}[\tilde{W}_2 W_3] \tilde{W}_4 W_1.
\end{eqnarray}
where $J[W] = 1_{\mathbb{C}^2} \, \mathrm{tr}[W] - W$. We again make a change of variables $k_2 \leftrightarrow k_3$, which implies that $\delta(k_1 - k_2 + k_3 - k_4)$ $\rightarrow$ $\delta(k_1 + k_2 - k_3 - k_4)$,
and results in the integrand
\begin{eqnarray}
\fl \tilde{W}_1 W_2 \mathcal{J}[\tilde{W}_3 W_4] 
	+ \mathcal{J}[W_4 \tilde{W}_3] W_2 \tilde{W}_1 
		- W_1 \tilde{W}_2 \mathcal{J}[W_3 \tilde{W}_4] - \mathcal{J}[\tilde{W}_4 W_3] \tilde{W}_2 W_1.
\end{eqnarray}

The conservative part can be written as a commutator
\begin{eqnarray}
\fl \mathcal{A}[W]_{1234} - \mathcal{A}[W]^*_{1234} \nonumber \\
	= [- W_3 + W_4 W_3 + W_4 \mathrm{tr}[W_2] + W_3 W_2 - W_4 W_2 - W_4 \mathrm{tr}[W_3], W_1].
\end{eqnarray}
In the second integral in (\ref{result}) we then exchange $k_2 \leftrightarrow k_3$, leading to $\delta(k_1 - k_2 + k_3 - k_4)$ $\rightarrow$ $\delta(k_1 + k_2 - k_3 - k_4)$ and resulting in the integrand 
\begin{eqnarray} \label{BminusB}
[- W_2 + W_4 W_2 + W_2 W_3 - W_4 W_3 + W_4 \mathrm{tr}[W_3] - W_4 \mathrm{tr}[W_2], W_1].
\end{eqnarray}
Hence the second term in (\ref{result}) is equal to $-\mathrm{i} t \langle \mathrm{f}, \,[H_{\mathrm{eff}}[W]_1,W_1]\mathrm{g} \rangle$ with $H_{\mathrm{eff}}[W]$ defined by (\ref{eq:Heff:1}).
On the other hand, using the symmetry of the delta-function under $k_3 \leftrightarrow k_4$, 
we can conclude that replacing (\ref{BminusB}) in the integrand by 
\begin{eqnarray}
\fl 
\frac{1}{2}[ W_3 \mathcal{J}[\tilde{W}_2 W_4] + \mathcal{J}[W_4 \tilde{W}_2] W_3 + \tilde{W}_3 \mathcal{J}[W_2 \tilde{W}_4] 
	+ \mathcal{J}[\tilde{W}_4 W_2] \tilde{W}_3, W_1 ] 
\end{eqnarray}
yields the same result.  

Therefore, in summary, the second order results are compatible with defining
\begin{equation}
\label{eq:BoltzmannEquation}
 \mathcal{C}[W](k,t)= \mathcal{C}_\mathrm{c}[W](k,t) + \mathcal{C}_\mathrm{d}[W](k,t),
\end{equation}
where
\begin{equation}
\label{eq:Cc}
\mathcal{C}_\mathrm{c}[W](k,t) = -\mathrm{i}\,[H_\mathrm{eff}(k,t), W(k,t)]
\end{equation}
and $H_{\mathrm{eff}}$ is defined either by (\ref{eq:Heff:1}) or by 
\begin{eqnarray}
\label{eq:Heff}
\fl 
H_{\mathrm{eff},1} = -\frac{1}{2} \int_{\mathbb{T}^d} \mathrm{d}k_2 \mathrm{d}k_3 \mathrm{d}k_4 \, \delta(k_1 + k_2 - k_3 - k_4) \, 
	\mathcal{P} \left(\case{1}{\omega_1 + \omega_2 - \omega_3 - \omega_4}\right) 
\nonumber \\
\times \big( W_3 \mathcal{J}[\tilde{W}_2 W_4] + \mathcal{J}[W_4 \tilde{W}_2] W_3 + \tilde{W}_3 \mathcal{J}[W_2 \tilde{W}_4] 
	+ \mathcal{J}[\tilde{W}_4 W_2] \tilde{W}_3 \big),
\end{eqnarray}
$\omega_i = \omega(k_i)$, $i \in \{ 1, 2, 3, 4 \}$, etc.  (The latter form was used as the starting point in \cite{LMS12}.)
\section*{References}




\end{document}